\documentclass[smallcondensed]{svjour3}

\expandafter\let\csname equation*\endcsname\relax
\expandafter\let\csname endequation*\endcsname\relax 
\usepackage{amsfonts}
\usepackage{amsmath}
\usepackage{amssymb}
\usepackage{amsbsy}
\usepackage{color}
\usepackage{graphicx}
\usepackage{url}
\usepackage{pgf}
\usepackage{tikz}
\usetikzlibrary{shapes,decorations,arrows}
\usepackage{cite}

\usepackage{authblk}

\usepackage{subcaption}
\usepackage{booktabs}

\usepackage{algorithm}
\usepackage{algorithmicx}
\usepackage{algpseudocode}

\newcommand{\ik}{$i${\bf k}}
\newcommand{\ad}{$ad$}

\newcommand{\PPPM}{PPPM}

\newcommand{\PD}{{\sc pppm}}

\newcommand{\ewald}{$\alpha$}
\newcommand{\rc}{$r_c$}
\newcommand{\p}{$p$}
\newcommand{\gs}{$(n_x \times n_y \times n_z)$}
\newcommand{\sunits}{$\sigma^{-1}$}

\newcommand{\tcube}{{\it TestLC}}
\newcommand{\tint}{{\it TestSI}}
\newcommand{\tbulk}{{\it TestSB}}

\begin{document}

\title{Accelerating scientific codes by\\performance and accuracy modeling}

%

\author{
Diego Fabregat-Traver \and Ahmed E. Ismail \and Paolo Bientinesi
}

\institute{
    Diego Fabregat-Traver \at 
        Aachen Institute for Advanced Study in Computational Engineering Science (AICES), RWTH Aachen,
        Schinkelstr. 2, 52062 Aachen, Germany.
        \email{fabregat@aices.rwth-aachen.de}
    \and
    Ahmed E. Ismail \at
        Department of Chemical Engineering,
        West Virginia University,
        PO Box 6102, Engineering Sciences Building,
        Morgantown, West Virginia 26506, USA.
        \email{ahmed.ismail@mail.wvu.edu}
    \and
    Paolo Bientinesi \at
        Aachen Institute for Advanced Study in Computational Engineering Science (AICES), RWTH Aachen,
        Schinkelstr. 2, 52062 Aachen, Germany.
        \email{pauldj@aices.rwth-aachen.de}
}

\maketitle

\begin{abstract}
Scientific software is often driven by multiple parameters that affect both
accuracy and performance.  Since finding the optimal configuration of these
parameters is a highly complex task, it extremely common that the software
is used suboptimally.  In a typical scenario, accuracy requirements are
imposed, and attained through suboptimal performance.  In this paper, we
present a methodology for the automatic selection of parameters for
simulation codes, and a corresponding prototype tool.  To be amenable to
our methodology, the target code must expose the parameters affecting
accuracy and performance, and there must be formulas available for error
bounds and computational complexity of the underlying methods.  As a case
study, we consider the particle-particle particle-mesh method (\PPPM{})
from the LAMMPS suite for molecular dynamics, and use our tool to identify
configurations of the input parameters that achieve a given accuracy in the
shortest execution time.  When compared with the configurations suggested
by expert users, the parameters selected by our tool yield reductions in
the time-to-solution ranging between 10\% and 60\%.  In other words, for
the typical scenario where a fixed number of core-hours are granted and
simulations of a fixed number of timesteps are to be run, usage of our tool
may allow up to twice as many simulations.  While we develop our ideas
using LAMMPS as computational framework and use the \PPPM{} method for
dispersion as case study, the methodology is general and valid for a range
of software tools and methods. 
\end{abstract}

\section{Introduction}

Simulation software  is often governed by a number of parameters that affect
both the accuracy of the results and the time-to-solution.  In a typical
setting, the choice of these values represents a challenging trade-off
scenario: the more accuracy is desired, the more computation is required (thus
longer simulations), and vice versa.
Users, who normally aim at a target accuracy level, 
face the problem of choosing $p$, 
a {\em configuration} of the parameters (i.e., a tuple of values), 
on a given set of computing resources,
that fulfills the accuracy requirements while minimizing
the execution time:
$$\min_{p} \ \text{time}(\text{resources}, p) \quad
\text{subject to} \quad \text{accurate}(p). $$
The problem is exacerbated by the large space of possibilities,
the intricate relation among the parameters, and the dependence
on the actual simulated system and underlying architecture.
In general, given the dimensionality of the space of configurations,
finding the optimal values for the parameters is a daunting task,
and even experts necessitate a considerable amount of trial and error 
to only provide rules of thumb that often are suboptimal.
Users are left with two options: either use the (potentially) suboptimal
rules of thumb from the literature, or perform a tedious and time
consuming search, which requires knowledge from the application domain,
the solver, and the underlying computing architecture.
In this paper, we present a methodology for the automatic parameter
selection for simulation codes, aiming at both an increase in productivity
and an improved utilization of computing resources of scientific simulations. 
A case study on one of the most popular methods in molecular dynamics
(the particle-particle particle-mesh method~\cite{PPPM-Hockney}) demonstrates the potential savings
offered by the methodology.

To be amenable to our methodology, a numerical code must 
present the following three characteristics. 
First, it has to be governed by a number of parameters that affect the efficiency
of the computation and/or the accuracy of the results; these parameters must be
exposed to the user, typically as input arguments to the simulation software.
Second, analytical formulas as functions of the parameters need to be available
for the estimation of the error incurred by a given configuration.
Finally, rough (asymptotic) cost estimates, generated either manually
or analytically, are required.
If these three requirements are satisfied, the methodology
proceeds in three steps.
\begin{enumerate}
    \item The first step consists in characterizing
the parameters that represent the search space;
this involves identifying those parameters that affect performance and/or
accuracy, choosing meaningful ranges for them, and discretizing the continuous ones. 
We refer to the set of all possible values for these parameters
as the ``parameter space'' or ``search space''.
    \item In the second step, analytical error bounds are used to divide 
the search space into accurate and inaccurate configurations, according to
whether or not they are estimated to satisfy the user
requirements; only the accurate subspace is further considered.
    \item As a third step, the execution time of the method is modeled by fitting 
one or more functions corresponding to the computational cost of the method to
data samplings (collected from short runs);
the combination of these functions yields a model that accurately predicts the
execution time for each configuration in the accurate subspace.
\end{enumerate}

The description of the steps of the methodology is deliberately general.  In
practice, their application will be adjusted to the properties of the method
to overcome the curse of dimensionality:  While the first stage only
requires acquiring a high-level understanding of the method or software at
hand, the second and third stages require actual computation, and the
potentially high dimensionality of the search space poses challenges to an
accurate prediction and selection of parameters.
To overcome these challenges, it is critical to exploit
method-specific properties and knowledge in order to reduce the complexity of the
search and to obtain faster and more accurate predictions.

To illustrate the potential of the methodology, we developed
a prototype that implements the methodology and applied it to 
the particle-particle particle-mesh (PPPM) method.
This method is governed by four parameters (three of them
affect performance, and all four affect accuracy) leading to
a large parameter space. 
Moreover, the overall accuracy of the simulation may be regulated by 
multiple accuracy thresholds, corresponding to different sections
of the method.
In general, in order to remove the need for a manual search of good configurations and to
simplify the user's workflow, the developers of the solvers provide rules of thumb
on how to set these parameters.
However, since the optimal choice highly depends on both the
actual simulation and the architecture,  the effectiveness
of these guidelines is limited.
In contrast, as we demonstrate, when both the simulated system and the computing architecture are
taken into account, it is possible to identify configurations that lead to
close-to-optimal performance, and thus to an efficient use of resources.

The benefits of our tool are two-fold.
On the one hand, it provides the users with close-to-optimal
configurations specifically chosen for the system and architecture
of interest.
On the other hand, it does so while dispensing them from the burden of an 
unpleasant and time-consuming manual search for such efficient configurations.
Moreover, the tool does not require any deep understanding of
the solvers or computing architectures. The user can thus focus
on the high-level aspects of the scientific problem.

In short, our experiments demonstrate how even expert choices for parameters might be
severely suboptimal in terms of efficiency: While the simulations deliver
the required accuracy, they do not do so in the most efficient way.  In other
words, resources are underutilized. 
At the expense of an initial (automated) search, our approach
yields gains in terms of productivity and better usage of resources.
This is especially relevant in the common cases where the simulations
take considerable time or many simulations with similar characteristics
are to be run.
More specifically, in our experiments we observed reductions
of time-to-solution between 10\% and 60\%. Given the length
of typical molecular dynamics simulations, this may translate to
savings of hours or even days of computation, or the execution
of twice as many simulations given a fixed core-hour budget.

\subsection{Contributions.} 

The main contribution of this paper is a methodology for the automatic
parameter selection for simulation software. The requirements for its
application are that the parameters affecting accuracy and performance of the
software are exposed to the user, and that formulas for error bounds and
computational complexity of the underlying solvers are available.
The outcome of the process is a configuration of the parameters that
yields accurate enough results in (almost) minimal time.
We focus on the domain of molecular dynamics, and contribute a practical example
of the potential benefits of our approach based on a very popular method in the field,
the particle-particle particle-mesh method (\PD{})~\cite{PPPM-Hockney}, and its implementation
from the well-known LAMMPS suite~\cite{LAMMPS-1995}.\footnote{For a list of papers citing LAMMPS,
many of which present results using this software, please visit
\url{http://lammps.sandia.gov/papers.html}.} 
Usage of our prototype implementing the methodology does not require deep knowledge
of the solvers and computing architectures, and at the cost of an easily
amortized automated search, the tool provides the user with close-to-optimal
configurations.
As a result, researchers are enabled to carry out many more or larger simulations
and therefore to gain deeper scientific insights in the problem at hand.

\subsection{Outline of the paper.}
This paper is structured as follows. Section 2 provides an overview of the basic
ideas behind molecular dynamics and the PPPM method.
Sections 3, 4 and 5 discuss in detail the three steps in our methodology, with
practical examples using the PPPM method. These steps are: characterization of
the search space, identification of the accurate subspace, and sampling and modeling.
In Sec. 6 we present multiple experimental results, while in Sec. 7 we draw conclusions. 

\section{Background}

This section reviews the basic ideas behind molecular dynamics and the PPPM
method, as well as research efforts related to the presented work.  The readers
familiar with both molecular dynamics and PPPM may skip this section.

\subsection{Molecular Dynamics and the PPPM method}

Molecular dynamics (MD)  is a well-established tool for
the study of the properties of complex particle systems at the atomistic level;
it is widely used in a variety of fields, including computational chemistry, biophysics, and 
materials science. 
Typical simulations consist of systems comprising between 
thousands and tens of millions of particles.  In order to simulate time scales
relevant for the processes being studied, and thus to obtain meaningful
evidence, these systems must evolve for at least millions of timesteps.
In practice, MD simulations are limited by computing resources,
and practitioners usually have to apply for compute time on supercomputers.
It is therefore critical to make an efficient use of the available resources.

The basic idea underlying an MD simulation is to study the movement of the
particles due to the forces acting on each of them, for a given time span.
The computation in these simulations is dominated 
by the calculation of the forces exerted on each particle 
(or, similarly, the calculation of the potential energy of the system).
Given a system with $n$ particles, the direct evaluation of pairwise
interactions would cost $O(n^2)$ operations (per timestep), and 
is thus entirely infeasible for systems with a large number of particles.
The so-called mesh-based Ewald methods, among which we find the PPPM method,
reduce the algorithmic complexity
to $O(n\,\,log \,n)$~\cite{PPPM-Hockney,PME-Darden,SPME-Essmann}.

In order to reduce the computational cost of one timestep from $O(n^2)$
to $O(n\,log\,n)$, \PPPM{} splits the the interactions
into short- and long-range ones. Forces among neighboring particles within a
given {\it cutoff} radius are computed in {\it real} space by means of direct evaluation,
while forces due to the interaction of distant particles are computed
in Fourier (or {\it reciprocal}) space. 
The calculation of the reciprocal space contributions, that is, the long-range interactions,
requires solving the Poisson equation in Fourier space. In order to take
advantage of the Fast-Fourier Transform (FFT) algorithm and achieve
the $O(n\,log\,n)$ complexity, the potential of
the particles is mapped into a regular grid, computations based on FFTs
are performed, and the resulting potential is mapped back to the particles.
Depending on the specifics of how the Poisson equation is solved, multiple
flavors of \PPPM{} arise. In the following, we consider two of them:
analytical differentiation (\ad{}), and \ik{} numerical
differentiation (\ik{}). For details on these two flavors
we refer the reader to~\cite{MeshUpI}.

A simulation based on the \PPPM{} method is governed by 4 parameters: 
the cutoff, which restricts the short-range contribution
 to particles within a certain radius;
the size of the grid into which the particles are mapped for the 
 calculations of the long-range interactions;
the interpolation order, which affects the mapping of potential into the grid,
 and indicates the number of grid points (per dimension)
 to which the potential is mapped; and
the Ewald parameter, which controls the weight
of each (short- and long-range) contribution.
Out of these four parameters, the first three (cutoff, grid size, and
interpolation order) affect both the accuracy and execution time of the simulation,
while the Ewald parameter affects the accuracy but not the execution time.

The impact of the cutoff, grid size, and interpolation order is rather
straightforward. When the cutoff is increased, more particles are taken into
consideration, the accuracy of the real space part also increases, and
the computation becomes more expensive.
Similarly, an increase in the grid size or in the interpolation order results
in higher accuracy and computational cost for the reciprocal space part.
The role of the Ewald parameter (\ewald{}) is more subtle.  While it does not
play a role in terms of computation, and thus performance, it has a strong
influence on accuracy. For instance, for a fixed cutoff, grid, and interpolation
order, larger values of \ewald{} improve the accuracy of the real space and
reduce that of the reciprocal space.
This fact can be used to optimize performance: Given a configuration that
attains the desired accuracy but is suboptimal in terms of performance,
the value of \ewald{} can be modified to shift the contribution
to compute time from one part to the other.

\begin{figure}
    \centering
    \includegraphics[scale=0.5]{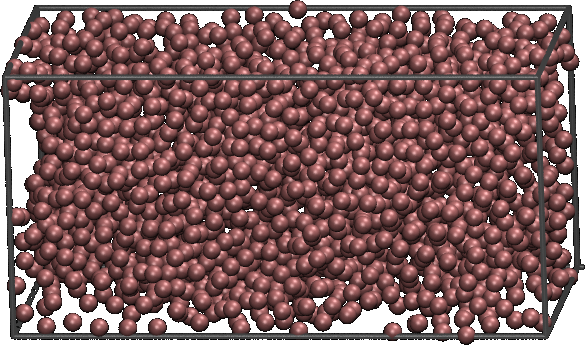} \hspace{2cm}
    \includegraphics[scale=0.25]{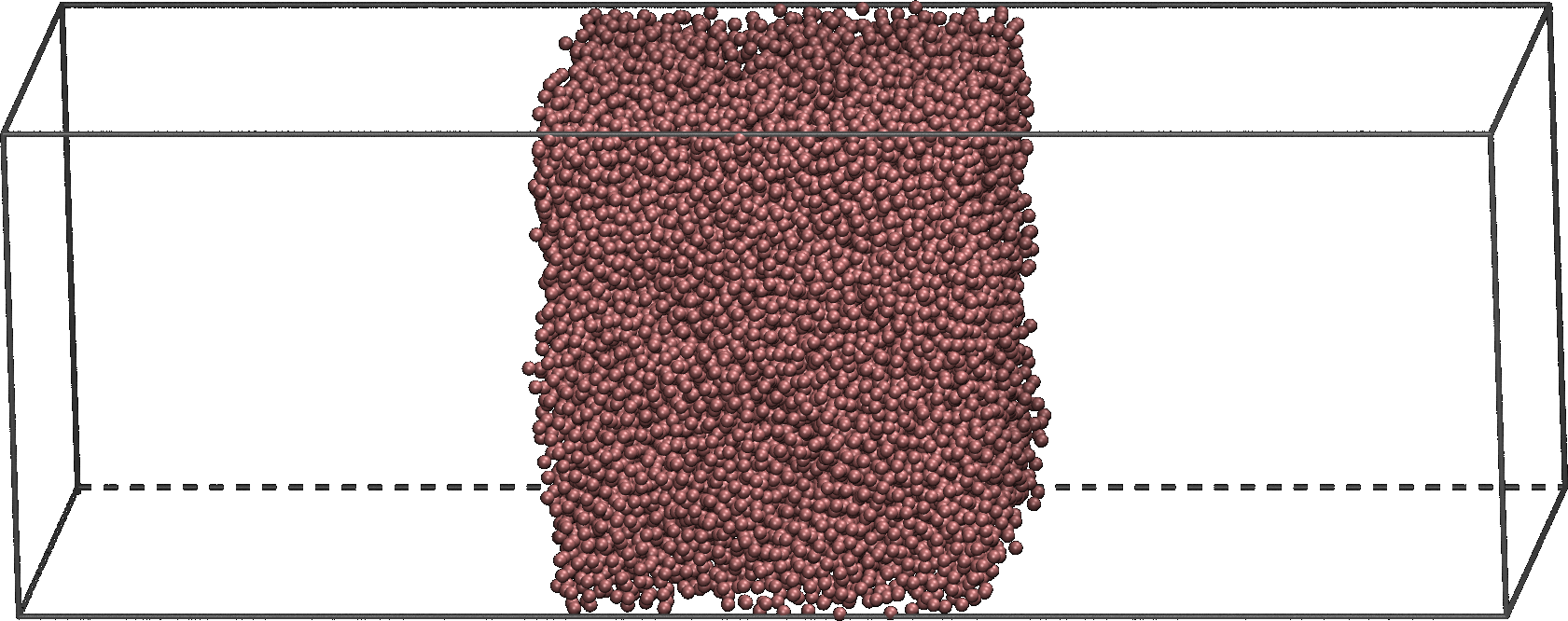}
    \caption{Two types of systems. Left, bulk system. Right, interfacial system.}
    \label{fig:systypes}
\end{figure}

To showcase our methodology, we choose the two types of systems
depicted in Fig.~\ref{fig:systypes}:
bulk (homogeneous) and interfacial (inhomogeneous). 
Homogeneous systems, with a random distribution of particles over the entire
domain, are typically used to initially test accuracy, and
 performance models. Inhomogeneous systems are very common and the most relevant in
practice; they constitute a
class of complex enough systems to stress the effectiveness of our methodology.

As a specific implementation of the PPPM solver, we choose the PPPM
solver for dispersion interactions from the LAMMPS package, a widely-used open
source MD suite.
The choice of the
solver for dispersion is not arbitrary; dispersion forces are
a type of forces that exist between all types of atoms and are
therefore present in every system. Of course, our approach is
applicable to other types of forces, such as the electrostatic ones.

Our tool takes as input a description of the simulation
to be run and the desired accuracy, and returns as output the estimated fastest
configuration that satisfies the accuracy constraints.
The input description includes the size of the simulation domain,
the number of particles in the domain, and whether they fill up the entire
domain (bulk), or only a box within the domain (interfacial).
The desired accuracy is expressed as either two independent thresholds
for the short- and long-range contributions ($\Delta F_{real}$ and 
$\Delta F_{reciprocal}$, respectively), or a single value as a threshold
for the combined root mean square
({\footnotesize $\sqrt{\Delta F_{real}^2 + \Delta F_{reciprocal}^2}$ }),
where $\Delta F_\star$ is defined as $\sqrt{\frac{1}{N} \sum_{i=1}^N
(F_i - F_i^{\text{exact}})^2}$, $N$ the number of particles.
The tool returns the estimated optimal values for cutoff, grid size,
interpolation order, and Ewald parameter.

\subsection{Related work}

Research efforts in the domain of molecular dynamics simulations concentrate
mainly in the design of accurate and efficient methods and their
parallel scalable implementation in software packages.
The MD landscape is populated with a broad variety of methods,
from simple truncation (with or without tail correction), through
tree methods, and grid-based methods. The latter group contains, among
others, the particle-mesh Ewald (PME), smooth particle-mesh Ewald (SPME), 
the particle-particle particle-mesh (\PPPM{}), and the multi-level
summation (MSM) methods. Our methodology is applicable to all these methods.

The list of available molecular dynamics suites is also large. Among others, it
is worth mentioning GROMACS, NAMD,
and CHARMM.~\cite{GROMACS-1995,CHARMM-2009,NAMD} While in our case
study we consider LAMMPS, the approach is generic and totally portable to any other
suite.

Literature on the accuracy of the different methods is 
abundant~\cite{Kolafa1992,Petersen1995,Deserno:1998wb,Hardy2006}.
Furthermore, there exists literature on the optimal choice of certain
parameters for accuracy. Among them, in~\cite{MeshUpII}, the authors
discuss the choice of optimal Ewald parameter given fixed cutoff, 
grid size, and interpolation order for the \PPPM{} method in
its \ik{} differentiation flavor. The authors of~\cite{Stern-optimal-ewald}
perform a similar study for both \ad{} and \ik{} differentiation in \PPPM{}.
However, despite the importance of making an efficient usage
of the available resources in order to target larger systems, 
the optimal choice of parameters in terms of performance
has received much less attention. An attempt that shares some
resemblance with our approach is given in~\cite{Wang-optimal-perf}.
They propose an analytical approach to finding optimal
values for all four parameters in SPME (the same four as in \PPPM{}).
However, we observe two limitations in the
study. First, the fact that their approach does not take
into consideration the actual hardware. The authors work
under the assumption that every flop (arithmetic operation)
has the same cost; due to caching effects and the cost of data movement, it is
well understood that an accurate
model requires taking into account that the cost of flops is not constant
across a piece of software.
Second, their numerical
results do not provide any reference to understand how close the
execution times are to the optimal. As later discussed in this paper,
we determine the region in the parameter space that potentially
contains the optimal configurations, and compare the results of
our tool with the best timings in that region.
The fact that we take into consideration the architecture, also allows
to identify close-to-optimal configurations across computing platforms.

\section{Characterization of the search space}
\label{sec:space}

The first step in our methodology for the automatic selection of parameters 
is to characterize the parameter space, that is, to identify the
set of parameters $\mathcal{P}$ that play a role in the accuracy and/or performance
of the target method. 

For most algorithms in computational science, the set $\mathcal{P}$ of input
parameters is a mixture of (potentially unbounded) discrete and continuous
values. For each of these parameters, a range must be specified and, for the
continuous ones, a discretization (not necessarily regular) provided. This
process originates the search space $\mathcal{S}$ of parameter configurations.
Without loss of generality, when there is freedom in the choice, the
granularity of the discretization and the considered ranges of values are set
based on the experience of practitioners and domain experts. 

The objective of our methodology is to find the configuration, 
that is, the one point in the (highly dimensional) space $\mathcal{S}$, that delivers 
the requested accuracy in the minimum time.

\vspace{3mm}
\noindent
{\bf Example: Characterizing $\mathcal{S}$ for the \PD{} method. }
The \PD{} method is parameterized by 
the cutoff radius (\rc{}),
the grid size (\gs{}),
the interpolation order (\p{}),
and the Ewald parameter (\ewald{}).
Out of the four parameters, the interpolation order and the grid size are
already discrete, while the Ewald parameter and the cutoff are continuous.
In the LAMMPS implementation of \PD{}, the accepted values for the interpolation order are integers from 2 to 
6.
The grid size is restricted to values that can be expressed as
multiples of only 2, 3, and 5 (e.g., a grid
of size $60 \times 60 \times 60$ is valid, but not a grid of size 
$66 \times 66 \times 66$). 
To constrain the (infinite) number of possible grid sizes,
we dynamically set an upper bound based on the system 
under consideration.
This upper bound is limited so that
only grids containing a number of grid points up to 8 times 
the number of particles in the system (2x per dimension) and with a
shape proportional to the simulation domain are allowed.
This bound is generous---the optimal grid size is typically far from
the largest allowed---and may be decreased to reduce the search time.

With respect to the continuous parameters,
the Ewald parameter must be strictly positive, 
and is typically not much larger than 1.0\sunits{};
we allow values in the range (0.0, 1.0].
As for the cutoff, no hard constraints are imposed, other than being strictly positive;
however, it is accepted that it takes at least a value of 2.0$\sigma$.
Regarding the upper bound, we allow rather large cutoffs 
up to 6.0$\sigma$.
For the discretization of the Ewald parameter and the cutoff, 
we choose a step size of 0.01\sunits{} and 0.1$\sigma$, respectively.
We recall that in both cases one can certainly explore a larger space of values;
the aforementioned bounds are flexible and 
the validity of our methodology and results are not affected by these choices.

This discretization leads to a 4-dimensional search space, where each
configuration consists of a 4-tuple (\ewald{}, \rc{}, \gs{}, \p{}).
As we discuss in the next section, the evaluation of
error estimates for all configurations in $\mathcal{S}$ is computationally too
expensive and thus unfeasible in practice due to the introduced overhead.
Furthermore, it is expensive to develop an accurate performance model that
takes the entirety of the search space into account.  Therefore we advocate for
an approach that exploits the structure of the target methods to reduce
the dimensionality of the search.
For the \PD{} method (and an entire class of similar methods), this includes 
(1) the fact that only accurate configurations are worth considering, and
(2) the study of both accuracy and performance can be split using
a divide-and-conquer strategy into the study of its components,
namely the real- and reciprocal-space contributions, 
which are then composed to provide a global solution.

\section{Identification of the accurate subspace}

In this first computational stage of our methodology, 
accuracy bounds
are used as a discriminant to restrict the search space to
only those configurations that result in simulations
accurate enough to merit the effort of performance modeling.
Therefore, the discretized parameter space
is split into {\em accurate} and {\em inaccurate subspaces}, 
$\mathcal{S_A}$ and $\mathcal{S_I}$ respectively,
and only the former is kept for further consideration.
We refer to the boundary between both subspaces as the
{\em frontier} ($\mathcal{F}$). The frontier is a Pareto
Efficient Frontier comprising the configurations that are
Pareto optimal, that is, configurations that, while
satisfying the accuracy constraints, cannot reduce the contribution to the
computational cost of any one of the parameters without increasing the
contribution of the others or without compromising the accuracy of the solution
(crossing the boundary accurate-inaccurate).  

To estimate the accuracy of each configuration of parameters, we 
require the availability of formulas for the error bounds. 
These are typically derived, and provided by the developer
of each method in the corresponding publication. 
For \PD{}, the error bounds are provided in~\cite{Rolf1}, and
consist of two formulas, for the real space and the 
the reciprocal space contributions, respectively.
We outline these formulas.
The error of the real space contribution is bound by
$$\Delta F_{real} = \frac{{C} \sqrt{\pi} \alpha^5}{\sqrt{N V r_c}} %
  \left( \frac{6}{r^6_c \alpha^6} + \frac{6}{r^4_c \alpha^4} + \frac{3}{r^2_c \alpha^2} + 1 \right) %
  e^{-r^2_c \alpha^2},$$
  where $C$ is the dispersion coefficient (dependent on the particles in the system),
  $N$ is the number of particles in the system, $V$ is the volume of the system,
and $\alpha$ and $r_c$ are the Ewald parameter and the cutoff, respectively.

The error for the reciprocal space contribution is bound by
  $$\Delta F_{reciprocal} = C\sqrt{\frac{Q}{NV}},$$
where 
$$
    Q = \frac{1}{V} \sum_{{\bf k}\in \mathbb{M}}
          \left\{ 
              \sum_{{\bf m} \in \mathbb{Z}^3} \left\vert {\bf \tilde{R}} \left( {\bf k} + \frac{2\pi}{h}{\bf m} \right) \right\vert^2
        - \frac{ 
                        \left\vert 
                            \tilde{\bf D}({\bf k}) \sum_{{\bf m} \in \mathbb{Z}^3} \tilde{U}^2 ({\bf k} + \frac{2\pi}{h}{\bf m})
                            \tilde{\bf R}^{*} ({\bf k} + \frac{2\pi}{h}{\bf m})
                        \right\vert^2
                   }
                   { \vert\tilde{\bf D}({\bf k})\vert^2 [\sum_{{\bf m} \in \mathbb{Z}^3} \tilde{U}^2 ({\bf k} + \frac{2\pi}{h}{\bf m})]^2 }
          \right\}.
      $$
\noindent
The details of the last formula are beyond the scope of this paper.
The main message is that it is not uncommon that error bounds are given by
means of complex formulas, whose evaluation might be computationally expensive
and even require numerical approximations.

To limit the cost of the evaluation of the formulas, one can make
use of the available knowledge on the specific method at hand.
For instance, in \PD{}, since separate formulas are provided for both the real-
and reciprocal-space contributions, and some of the parameters affect only one
of the two errors, we decompose the evaluation in that of each space, and then
combine the results to obtain the error estimates for each configuration of the
four-dimensional space.  This approach is general and valid for a class of
methods with similar characteristics.

\begin{figure}[!h]
    \centering
	\begin{subfigure}[b]{0.44\textwidth}
		\centering
		\includegraphics[width=0.7\textwidth]{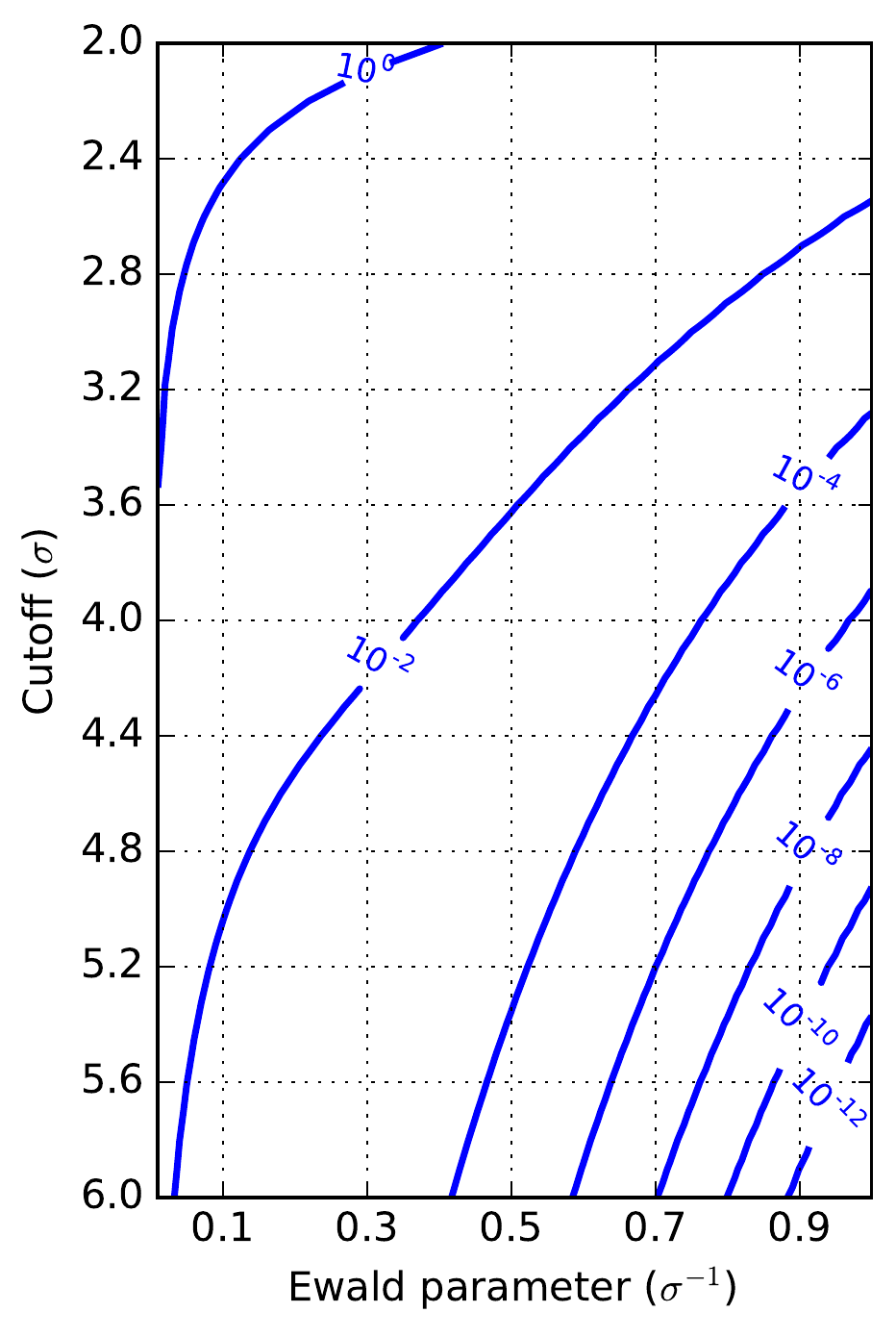}
		\caption{Real-space error.}
		\label{fig:acc-realspace}
    \end{subfigure}
	\hfill
	\begin{subfigure}[b]{0.50\textwidth}
		\centering
		\includegraphics[width=0.8\textwidth]{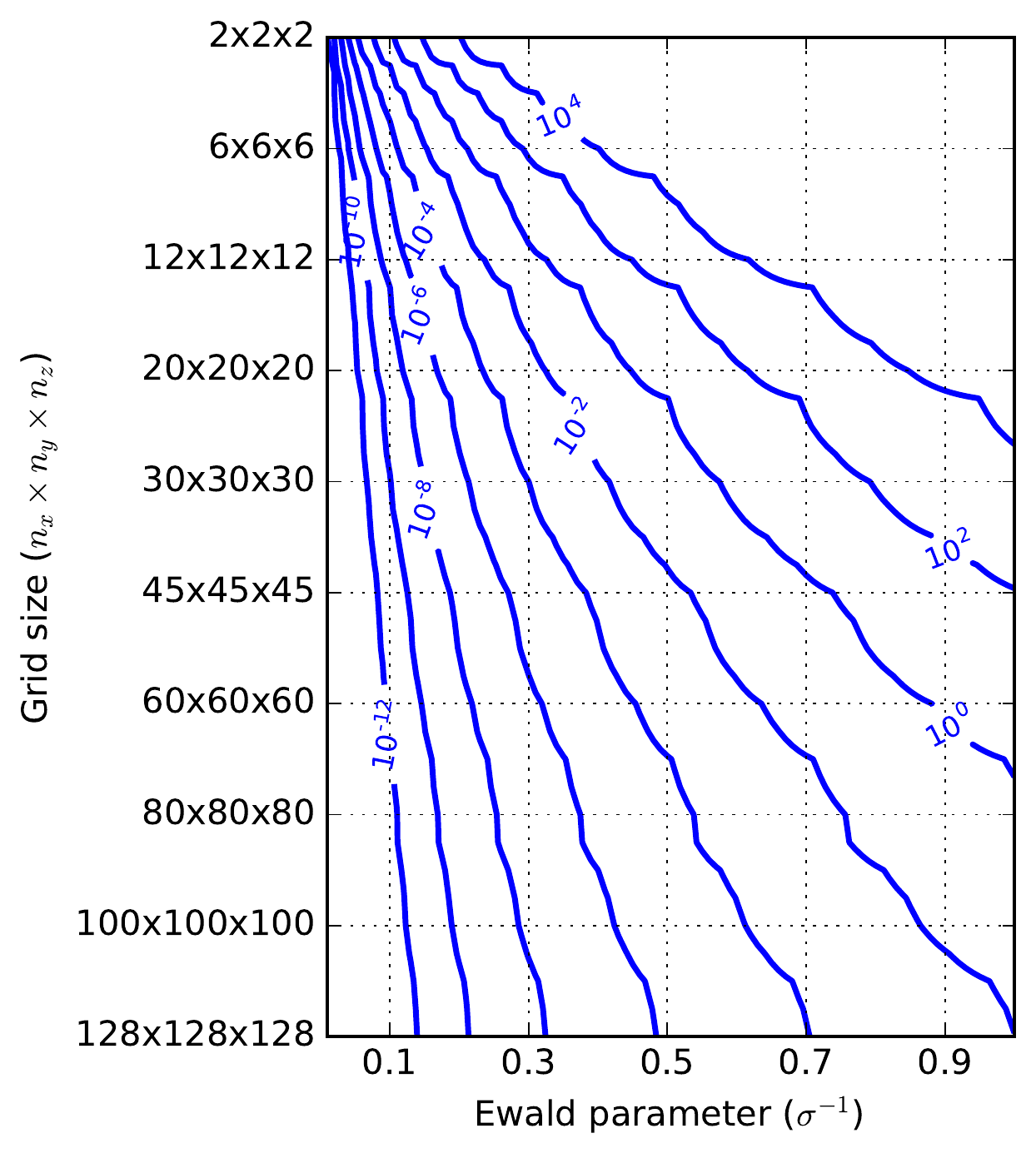}
		\caption{Reciprocal-space error.}
		\label{fig:acc-kspace}
	\end{subfigure}
    \caption{Contour plots for the study of the accurate subspace. 
    The plots correspond to \tcube{} using \ad{} differentiation. In the case of the reciprocal
        space, we show the error for an interpolation order of 5.} 
    \label{fig:acc-split}
\end{figure}

In \PD{}, the real space error is only affected by the choice of 
\ewald{} and \rc{},
while that of the reciprocal space 
is only affected by the choice of \ewald{}, \p{}, and \gs{}.
Figures~\ref{fig:acc-realspace} and~\ref{fig:acc-kspace} 
show, respectively, the independent evaluation of the error
estimates for the real and reciprocal spaces. The figures correspond
to the {\em Large Cube} test case (\tcube{}) described in Appendix~\ref{app:scenarios}.
The figures illustrate the tradeoffs and difficulties associated with 
the manual selection of parameters for a simulation run.
While higher values of the Ewald parameter
increase the real-space accuracy, they reduce the accuracy
of the reciprocal space. 
Also, for a fixed target accuracy, smaller values of the Ewald parameter allow
to use a smaller grid size or interpolation order, and hence to reduce the execution time for the
reciprocal-space calculation, at the expense of setting a larger cutoff, thus
increasing the execution time for the real space, and vice versa.
It becomes apparent that identifying the accurate subspace $\mathcal{S_A}$, and then determining
the fastest configurations within $\mathcal{S_A}$
is a daunting task.

While the evaluation of the real-space error formula is inexpensive,
the evaluation of the reciprocal-space error formula is, unfortunately, still too costly, even when
only the 3D subspace \ewald{} $\times$ \gs{} $\times$ \p{}
is considered. In fact, the values for Fig.~\ref{fig:acc-kspace} 
were calculated by an approximation also provided in~\cite{Rolf1}
which is only valid for the \ik{} differentiation, cubic domains, and
grid sizes with equal number of grid points in each dimension.
Without this simplification, the evaluation of the entire grid
using the generalized formula would take days.
We opt for an alternative that further reduces the amount of required
evaluations of the formula. The insight is that it
suffices to identify the values that split accurate and inaccurate configurations.
That is, referring to Fig.~\ref{fig:acc-kspace}, if the user requests an 
accuracy of $10^{-4} \epsilon/\sigma$, it suffices to find the corresponding contour plot;
every point below that line is accurate enough.
To this end, for each interpolation order and grid size,
we perform a binary search for this splitting value. Additionally, the search
is parallelized and carried out in place, making use of the same architecture in which 
the simulation will be run. 
Following this idea, 
the time spent in the evaluation of the error estimates may be reduced
from days to minutes.

Once the error estimates for both real and reciprocal spaces are
available, these are combined back in the single four-dimensional space $S$;
it is then possible to split the full search space $\mathcal{S}$
into $S_A$ and $S_I$,
according to the target accuracy.
The splitting of the search space is carried out in one of two different ways,
depending on how the user expresses the accuracy requirements.
If the user inputs one single value for the accuracy, 
those configurations where 
the combined root mean square of the errors
({\footnotesize $\sqrt{\Delta F_{real}^2 + \Delta F_{reciprocal}^2}$ })
is below the accuracy threshold are considered. 
If, instead, the user provides individual thresholds for each component,
then the individual errors must satisfy the corresponding constraint independently.

\begin{figure}[!h]
\centering
    \includegraphics[width=.6\textwidth]{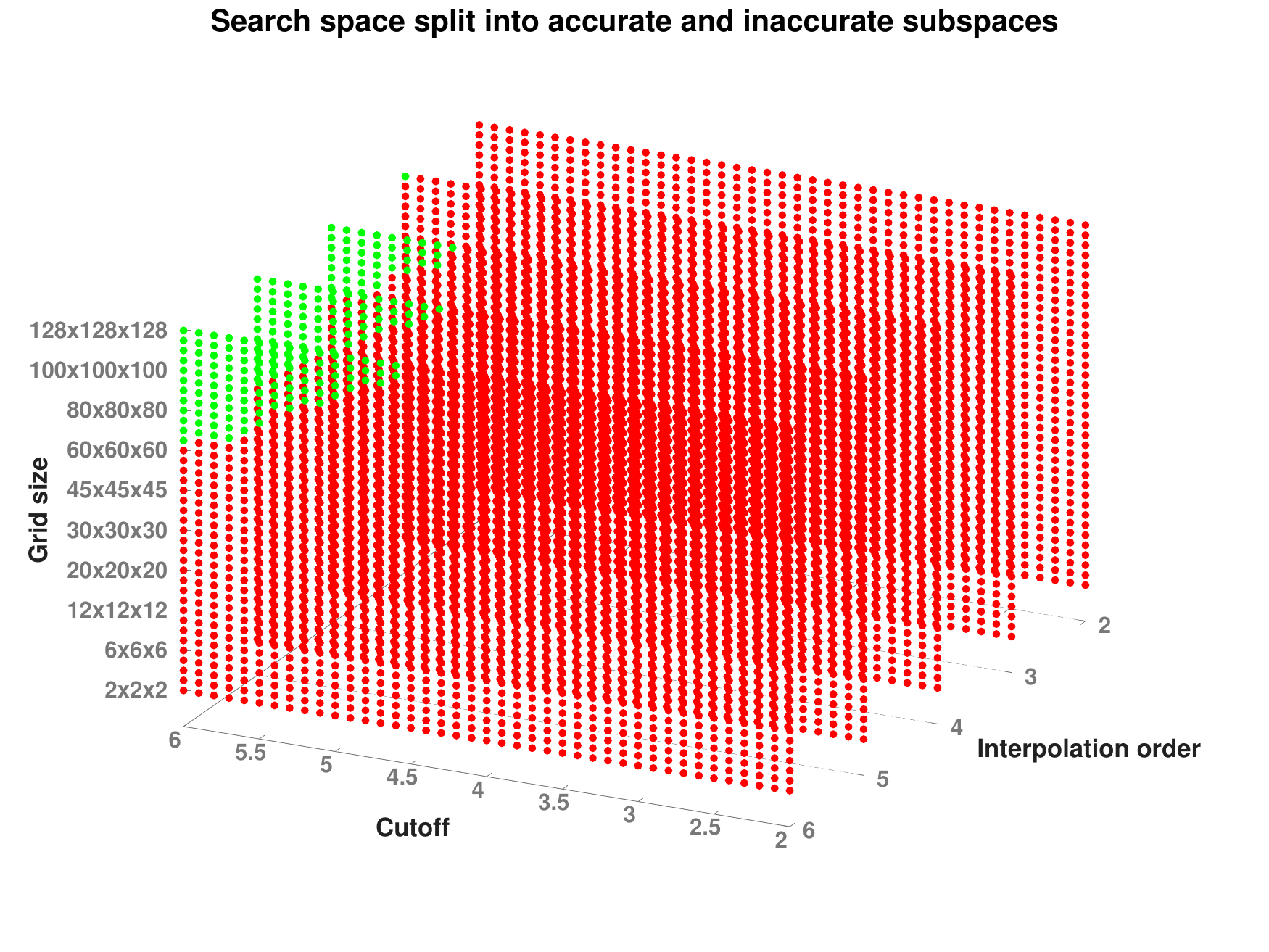}
    \caption{Search space divided into accurate and inaccurate configurations
        for \tcube{}.}
\label{fig:acc-space-ex}
\end{figure}

\begin{figure}[!h]
\centering
    \includegraphics[width=.6\textwidth]{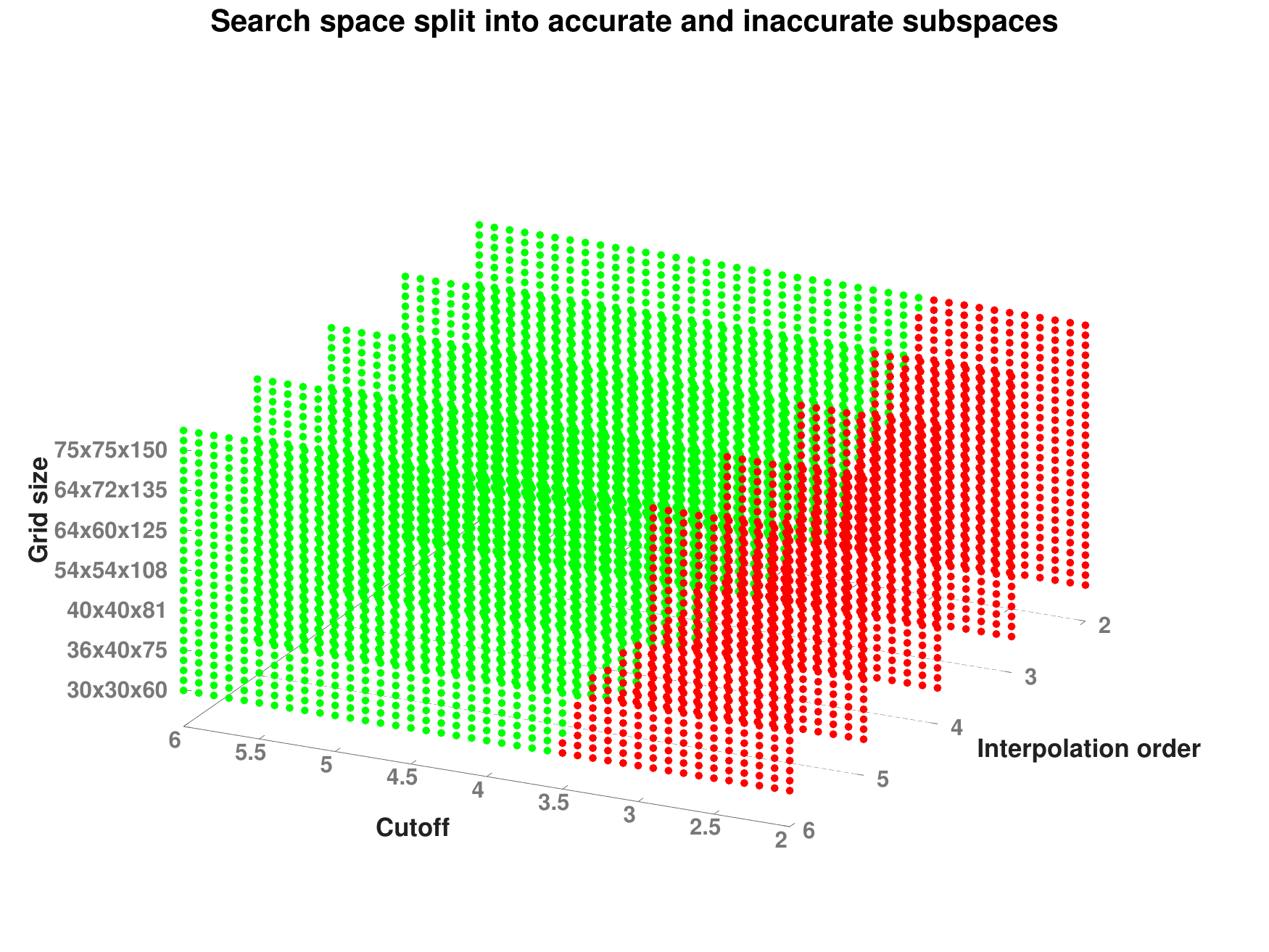}
    \caption{Search space divided into accurate and inaccurate configurations
    for the bulk system described in Sec.~\ref{subsec:exp-bulk}.}
\label{fig:csb-acc-space-ex}
\end{figure}

It is now important to point out that not all of 
the parameters that influence accuracy also impact performance. 
For instance, \ewald{} only affects accuracy, and
does not directly influence the amount of computation performed.
Thus, the space of parameters can be reduced to 3-dimensional 
(\rc{}, \gs{}, \p{}) when modeling performance.  At this
stage, points in this three-dimensional space are labeled as inaccurate, unless
there exists at least one value of \ewald{} that makes the combination
accurate.
Figures~\ref{fig:acc-space-ex} and~\ref{fig:csb-acc-space-ex}
illustrate this subdivision, respectively
for \tcube{}, and the bulk system used in our
experimental results (Sec.~\ref{sec:experiments}).
In both figures, green and red dots denote the accurate and inaccurate
subspaces, respectively.

\section{Sampling and modeling}
\label{sec:modeling}

Once the accurate subspace $\mathcal{S_A}$ has been identified,
the third and final
step consists in determining the configurations in $\mathcal{S_A}$ that lead to the best overall performance.
A range of alternatives exist to estimate the execution time given a choice of parameters.
At one extreme, one could rely on a purely analytical approach 
which consists of using the flop count
(the amount of arithmetic operations required by a given
configuration) as a metric to estimate the execution time.
Unfortunately, it is well-understood that not every flop incurs
the same cost: Different sections of a given implementation 
attain different performance levels
(flops/s), and even variations of the values for the parameters
influence the use of the memory hierarchy (reuse
of data), and thus performance.
Therefore, even though flops are a metric that allows for a
fast estimation of execution time, they lead to inaccurate predictions.
At the other extreme, one may consider a purely empirical approach, where each
of the accurate configurations is executed and timed, and the fastest one is
selected.  However, the number of configurations in $\mathcal{S_A}$ may vary
from several hundred to tens of thousands, depending on the particular system
and the desired accuracy. 
Such an extensive search would still consume
considerable resources. Thus, this approach is reliable, but slow.
Instead, we advocate a hybrid analytical-empirical approach, 
based on a reduced number of samples,
which predicts the performance for each configuration by fitting
a set of simple models to the measurements.

In Sec.~\ref{sec:static}, we present a static strategy that samples at
a set of predetermined points in the space. While rather effective and useful
to better understand the properties of the method, it presents some drawbacks
that we discuss in Sec.~\ref{sec:pros-cons}. In Sec.~\ref{subsec:adaptive-sampling}, we switch 
to a dynamic sampling strategy that reduces significantly 
the amount of required sampling and improves the accuracy of the estimations.
As our experiments confirm, the adaptive approach yields fast and reliable predictions.

\subsection{Static dense sampling}
\label{sec:static}

Our static approach consists in collecting samples at a set of predefined
points that cover a relatively dense subset of the space. The granularity
of the sampling obviously affects the accuracy and speed of the prediction:
the larger the number of samples, the slower and the more accurate is the
prediction, and vice versa.
If the space and the cost of sampling are too large, properties of the
method may be exploited to, for instance, reduce the dimensionality of the
space or speedup the samples.

For the \PD{} method, we take advantage of the fact that the method consists of two successive,
independent stages, and model each stage in isolation. The resulting models are
then combined to produce the predictions.
Similarly to the approach taken to evaluate the analytical formulas for the error bounds,
a divide-and-conquer strategy is used to decompose the performance modeling
of the \PD{} method into that of the real and reciprocal contributions.
Thus, the real and reciprocal spaces are sampled independently, the samples are then fitted 
to models, and finally the models are combined to predict the total compute
time of the accurate configurations.

\subsubsection{Sampling the real-space computation}
The only parameter that affects the time
required to compute the real-space contribution is the 
cutoff.
Hence, different values for the cutoff are sampled,
while the other parameters are fixed. More specifically,
we use the following configurations:
\begin{itemize}
    \item {Ewald parameter:} 0.50\sunits{}
    \item {Interpolation order:} 2
    \item {Grid size:} $1 \times 1 \times 1$ 
    \item {Cutoff:} [2.0$\sigma$, 2.5$\sigma$, 3.0$\sigma$, ..., 6.0$\sigma$]
\end{itemize}
Here, the choice of Ewald parameter is arbitrary, and the interpolation
order and grid size are set to the smallest possible value to
minimize the time spent in the sampling. We sample at multiple
values of the cutoff in the range $[2.0\sigma,\ 6.0\sigma]$ in steps of 0.5$\sigma$,
for a total of nine data points.

\subsubsection{Sampling the reciprocal-space computation}
The time spent in computing the reciprocal-space contribution is, in principle,
affected only by the grid size and the interpolation order. Hence, 
the cutoff and Ewald parameter are fixed, and the following configurations are sampled:
\begin{itemize}
    \item {Ewald parameter:} 0.50\sunits{}
    \item {Interpolation order:} [2, 3, 4, 5, 6]
    \item {Grid sizes:} a function of the target system (domain size and number of particles)
    \item {Cutoff:} 2.0$\sigma$.
\end{itemize}
Once again, the choice of the Ewald parameter value is arbitrary, the
cutoff is kept to a minimum, and we sample all interpolation orders in the
range $[2, 6]$, and the full set of valid grid sizes within a range determined
according to the number of particles and domain size of the target system. The
total number of data points varies with the size of the system, and is equal to five
(interpolation orders) times the number of considered grid sizes.

\vspace{5mm}
Here, and for the remainder of the paper, each sampling consists in the
execution of 1000 timesteps of the actual simulation of interest; of course,
this number is configurable. The rest of the properties of the simulation, such
as mixing rules, ensemble, temperature, etc, are fixed and configured by the user.

\subsubsection{Modeling and fitting}

\begin{figure}[!h]
  \centering
  \includegraphics[width=0.6\textwidth]{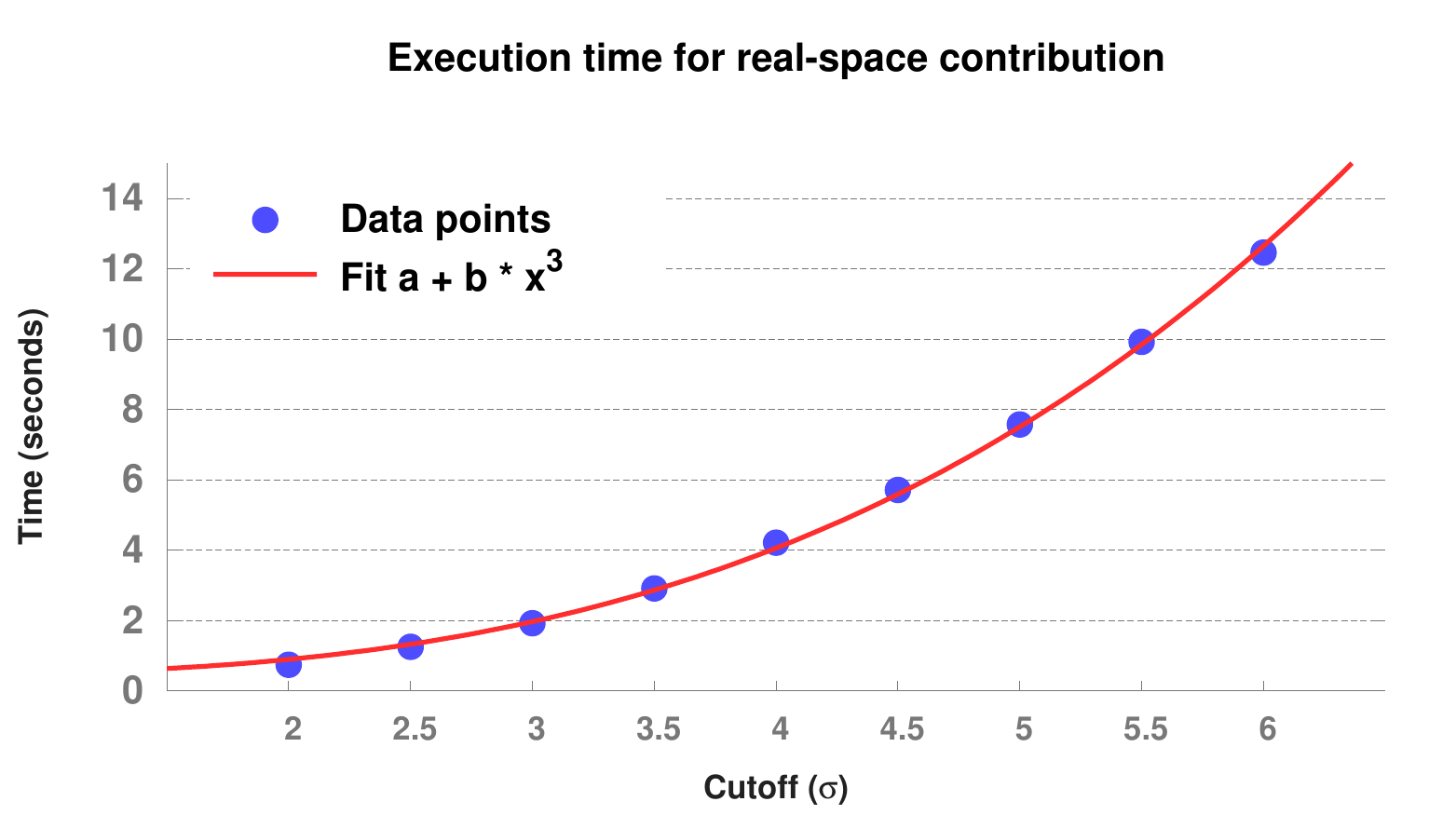}
  \caption{Fitting of the function $f(c) = a + b*c^3$ to the timings for the 
      real-space contribution in \tint{} (\ik{} differentiation).
    Values for parameters $a$ and $b$ are 0.44 and 0.0565, respectively.
    The average relative error is less than 5\%.}
  \label{fig:cutoff-full-sampling}
\end{figure}

Each set of samples is now fitted to a corresponding function.
The choice of the functions to be used comes either from domain expertise
or from the analysis of the method's complexity.
Since the computational cost for the evaluation of the real-space contribution
is proportional to the cube of the cutoff, we fit the function $f(c) = a + b\cdot c^3$
to the data points ($c_i$, $t(c_i)$), where the parameter $a$ accounts for the
overhead in allocation and setting of data.
As an example, we consider the test case {\em Small Interface} (\tint{}, see App.~\ref{app:scenarios}
for details).
Figure~\ref{fig:cutoff-full-sampling} shows the measured execution time for the
real-space contribution for each of the sampled cutoff values.
The fit is satisfactory, presenting an average relative error
$\frac{1}{n} \sum_{i=1}^n (|f(c_i)-t(c_i)|/t(c_i))$ of less than 5\%.

\begin{figure}[!h]
  \centering
  \includegraphics[width=0.6\textwidth]{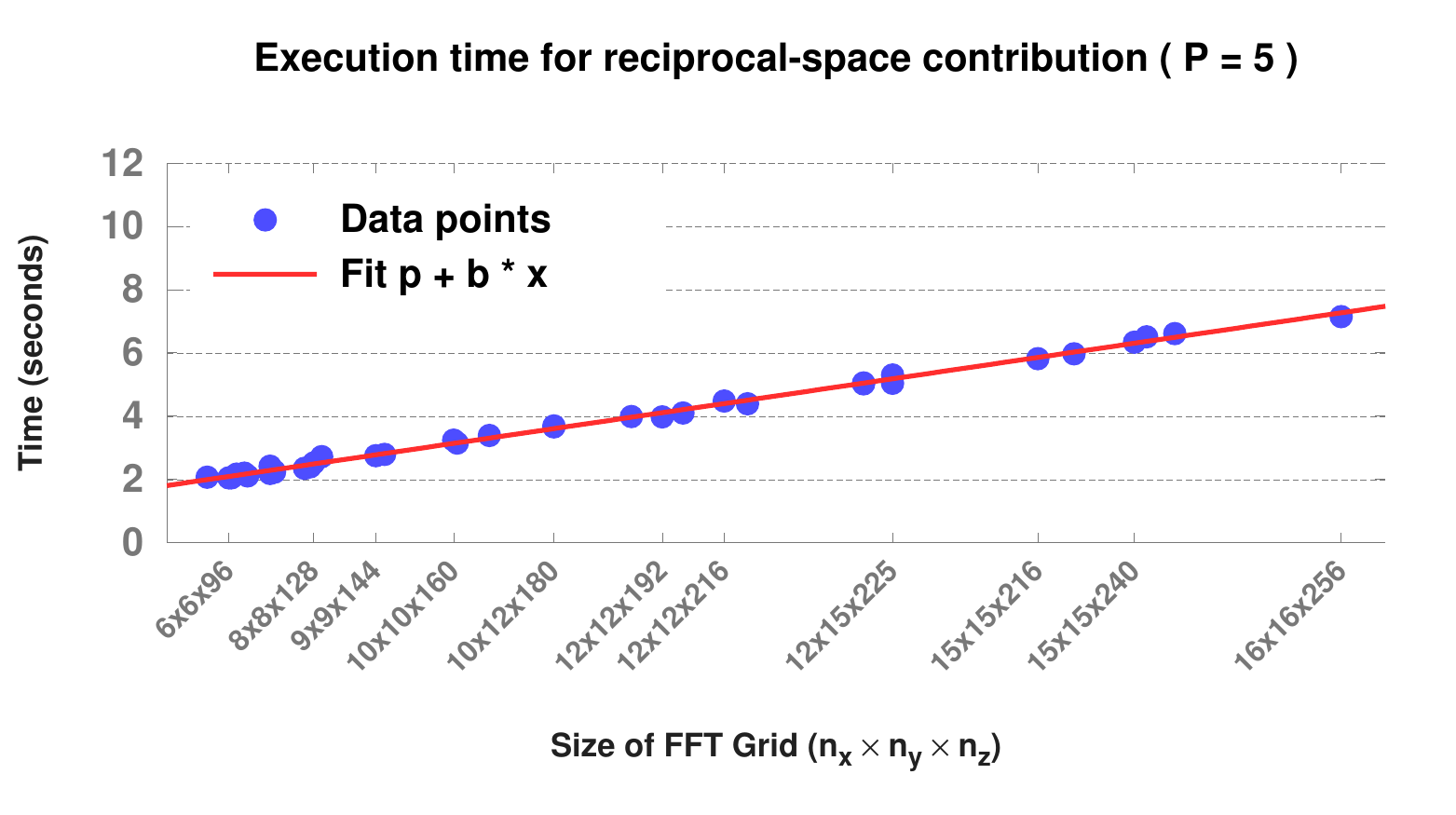}
  \caption{Fitting of the function $h_5(g) = p + b*g$ to the timings for the 
      reciprocal-space contribution in \tint{} (\ik{} differentiation).
  Values for parameters $p$ and $b$ are 1.81 and $8.33\cdot10^{-5}$, respectively.
    The average relative error is 2.3\%.}
  \label{fig:kspace-full-sampling}
\end{figure}

To simplify the modeling of the reciprocal space, we consider each
interpolation order $P$ independently.  Accordingly, we model the execution
time of the reciprocal space by means of multiple functions $h_i(g) = p + b \cdot g$, where
$g$ represents the number of points in the FFT grid, and $p$ accounts for the
time spent in the mapping of the particles into the grid and back.\footnote{
Even though the computational cost of the FFT is $O(g \; log(g))$, $g$ the number
of points in the grid, the empirical timings show that the implementation has a linear
cost. This comes at no surprise, since the implementation is communication-bound,
especially for large number of nodes.
}
Figure~\ref{fig:kspace-full-sampling} depicts the execution time for the
reciprocal space corresponding to $P=5$ and a range of grid sizes,
also for \tint{}. 
The average relative error is around 2\%.

\subsubsection{Prediction}

An estimate of the overall compute time is obtained by summing up the
estimates for the real-space and the reciprocal-space contributions.
We use test case \tint{} to provide preliminary results 
on the accuracy
of the predictions yielded by the described static approach. 
As a reference, we measured the time spent in the computation of the real and reciprocal 
space contributions for the frontier configurations and compared the timings
with the predictions for these same points.
Table~\ref{tab:full-sampling} collects the results for the fastest
five configurations based on empirical measurements (column 2).
Columns 3 and 4 show the predicted execution time for those same
configurations and the relative error in the prediction, measured as
$|t_{pred}-t_{emp}|/t_{pred}$, where $t_{emp}$ is the empirically measured
time and $t_{pred}$ is the predicted time.
The model-based predictions are quite accurate. The average relative error for
the entire set of frontier configurations (73) is 4.98\%.  Moreover, and most
importantly, the methodology correctly identifies the configuration which leads
to the fastest execution.

  \begin{table}[!h]
      \centering
      \begin{tabular}{c @{\hspace*{1em}} c @{\hspace*{1em}}|@{\hspace*{1em}} c c @{\hspace*{1em}}|| 
         @{\hspace*{1em}} c @{\hspace*{1em}}c @{\hspace*{1em}}c @{\hspace*{1em}}c} \toprule
          \multicolumn{2}{c| @{\hspace*{1em}}}{\bf Empirical} & 
          \multicolumn{2}{c||@{\hspace*{1em}}}{\bf Prediction} &
          \multicolumn{4}{c}{\bf Configuration} \\ \midrule
          Ranking & Time & Time & Error & \ewald{} & \rc{} & \p{} & \gs{} \\ \midrule
          \#1 & 8.770s & 8.378s & (-4.68\%) & 0.56\sunits{} & 4.60$\sigma$ & 4 & $10 \times 10 \times 160$ \\
          \#2 & 8.920s & 8.495s & (-4.99\%) & 0.60\sunits{} & 4.40$\sigma$ & 4 & $12 \times 12 \times 180$ \\ 
          \#3 & 9.023s & 8.595s & (-4.98\%) & 0.63\sunits{} & 4.30$\sigma$ & 4 & $12 \times 12 \times 216$ \\
          \#4 & 9.119s & 8.754s & (-4.17\%) & 0.58\sunits{} & 4.50$\sigma$ & 5 & $10 \times 10 \times 160$ \\
          \#5 & 9.299s & 8.610s & (-8.00\%) & 0.64\sunits{} & 4.20$\sigma$ & 5 & $12 \times 12 \times 180$ \\ \bottomrule
      \end{tabular}
      \caption{Predictions for \tint{} (\ik{} differentiation) using
          static sampling.}
      \label{tab:full-sampling}
  \end{table}

\subsection{Advantages and disadvantages of the static sampling}
\label{sec:pros-cons}

We observed a number of advantages and disadvantages of using
a static sampling. Among the advantages, we highlight the
simplicity of implementation, since the approach is system-agnostic,
that is, the system does not influence the search beyond the
selection of grid sizes to consider, and no online decisions
are necessary. Second, the accuracy of the predictions is rather
satisfactory in general. Finally, it is relevant beyond the automatic
selection of parameters. The sampling allows to understand the
behavior of the method in practice and to expose unexpected
performance signatures. We give examples below.

On the contrary, we also identified a number of drawbacks that
may imply limited accuracy in the predictions or excessive sampling time:

\begin{enumerate}
  \item Unexpectedly, the value of the cutoff does affect the execution time
      of the reciprocal-space computation. 
    \item Unlike in Fig.~\ref{fig:kspace-full-sampling}, in certain cases 
      the timings for the reciprocal space may not fit a single straight line;
      two shifted lines are observed instead (see Fig.~\ref{fig:kspace-shift} below).
  \item The number of required samples may grow large.
\end{enumerate}
These issues are discussed in detail hereafter. 
Our proposed solution, based on adaptive sampling, is developed in the next subsection.

\paragraph{Impact of \rc{} on the reciprocal space. }

\begin{figure}[!h]
  \centering
  \includegraphics[width=0.6\textwidth]{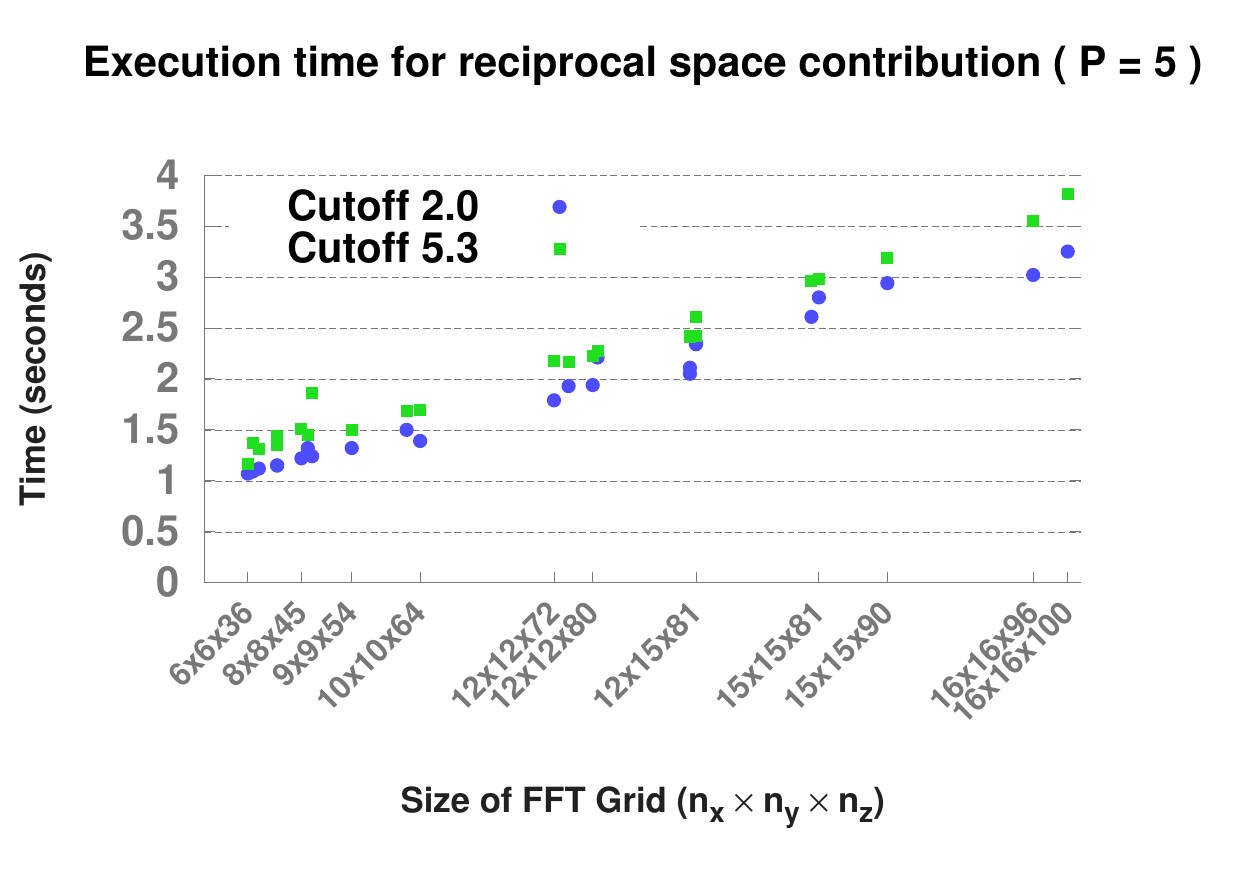}
  \caption{Difference in execution time for the calculation of the
  reciprocal-space contribution when using different cutoffs.}
  \label{fig:kspace-cutoff-effect}
\end{figure}

While, in principle, the cutoff should only have an impact on 
the execution time of the real-space contribution, we observed that the execution time of
the reciprocal space is also affected. 
This behavior is observed, for instance,
in the test case {\em Small Bulk} (\tbulk{}, see App.~\ref{app:scenarios}).
As illustrated by Fig.~\ref{fig:kspace-cutoff-effect}, the difference in
execution time when calculating the reciprocal-space contribution with
two different fixed cutoffs (in this case 2.0$\sigma$ and 5.3$\sigma$) may be considerable.
Indeed, these differences are carried on to the prediction of the execution
time of the overall simulation, as illustrated in Tab.~\ref{tab:cutoff-kspace}.
Columns 3 and 4 correspond to predictions after using a cutoff of 2.0$\sigma$
for the samplings of the reciprocal-space. As one can appreciate, the predictions
may be quite off. In fact, the average relative error between the measured and the 
estimated execution times when using this cutoff is about 5\%. 
If, instead, we take into account the range of cutoff values represented in the 
configurations included in $S_A$, and choose to sample using a value for the
cutoff within that range, the overall estimation improves. In the case of \tbulk{}, 
the cutoff in the frontier configurations ranges from 4.6$\sigma$ to 6.0$\sigma$. 
If we fix the cutoff for sampling to the mid value (5.3$\sigma$),
the average relative error
(Tab.~\ref{tab:cutoff-kspace}, columns 5 and 6) is reduced to about 2\%.
%
%
We thus conclude that is critical to dynamically choose the sampling values based
on the simulation under consideration in order to obtain highly-accurate predictions.

\begin{table}[!h]
  \centering
  \begin{tabular}{c @{\hspace*{1em}} c @{\hspace*{1em}} | 
                  @{\hspace*{1em}} c @{\hspace*{-4em}} c | 
                  @{\hspace*{1em}} c @{\hspace*{-4em}} c} \toprule
      \multicolumn{2}{c|@{\hspace*{1em}}}{\bf Empirical} & 
                  \multicolumn{2}{c@{\hspace*{1em}}|@{\hspace*{1em}}}{\bf Prediction (\rc{} = 2.0$\sigma$)} &
                  \multicolumn{2}{c@{\hspace*{1em}} }{\bf Prediction (\rc{} = 5.3$\sigma$)} \\ \midrule
      Ranking & Time & Time & Error & Time & Error \\ \midrule
      \#1 & 7.498s & 7.413s & (-1.15\%) & 7.679s & (+2.36\%) \\
      \#2 & 7.542s & 7.394s & (-2.01\%) & 7.687s & (+1.88\%) \\
      \#3 & 7.595s & 7.270s & (-4.47\%) & 7.610s & (+0.20\%) \\
      \#4 & 7.660s & 7.520s & (-1.86\%) & 7.759s & (+1.28\%) \\
      \#5 & 7.800s & 7.255s & (-7.52\%) & 7.736s & (-0.83\%) \\ \bottomrule
  \end{tabular}
  \caption{Results for \tbulk{} (\ik{} differentiation). The accuracy of the
      predictions improve when using a cutoff closer to the range in the
      frontier configurations ([4.6$\sigma$, 6.0$\sigma$]).}
  \label{tab:cutoff-kspace}
\end{table}

\paragraph{Irregular behavior of the reciprocal space.}

In some test cases, we observed that the timings for the reciprocal
space do not lay on one single line, but rather on 
two parallel ones (in a piecewise manner). We relate this behavior
to a switch in the data distribution in \PD{}'s implementation~\cite{Rolf1,Rolf2},
where depending on the grid size and the number of processes used for the computation,
the FFT domain is decomposed either in slabs (faces) or pencils (columns).
More specifically, the shift occurs at the point where the number of grid points in
the $z$ dimension becomes equal or greater than the number of processes used in
the simulation. As an example, Fig.~\ref{fig:kspace-shift} illustrates
the shift for the \tcube{} scenario. In the example, 96 processes
where used; the shift thus occurs at grid size $96\times96\times96$.
An adaptive sampling approach is required to identify the gap
and correctly fit the data.

\begin{figure}[!h]
  \centering
  \includegraphics[width=0.6\textwidth]{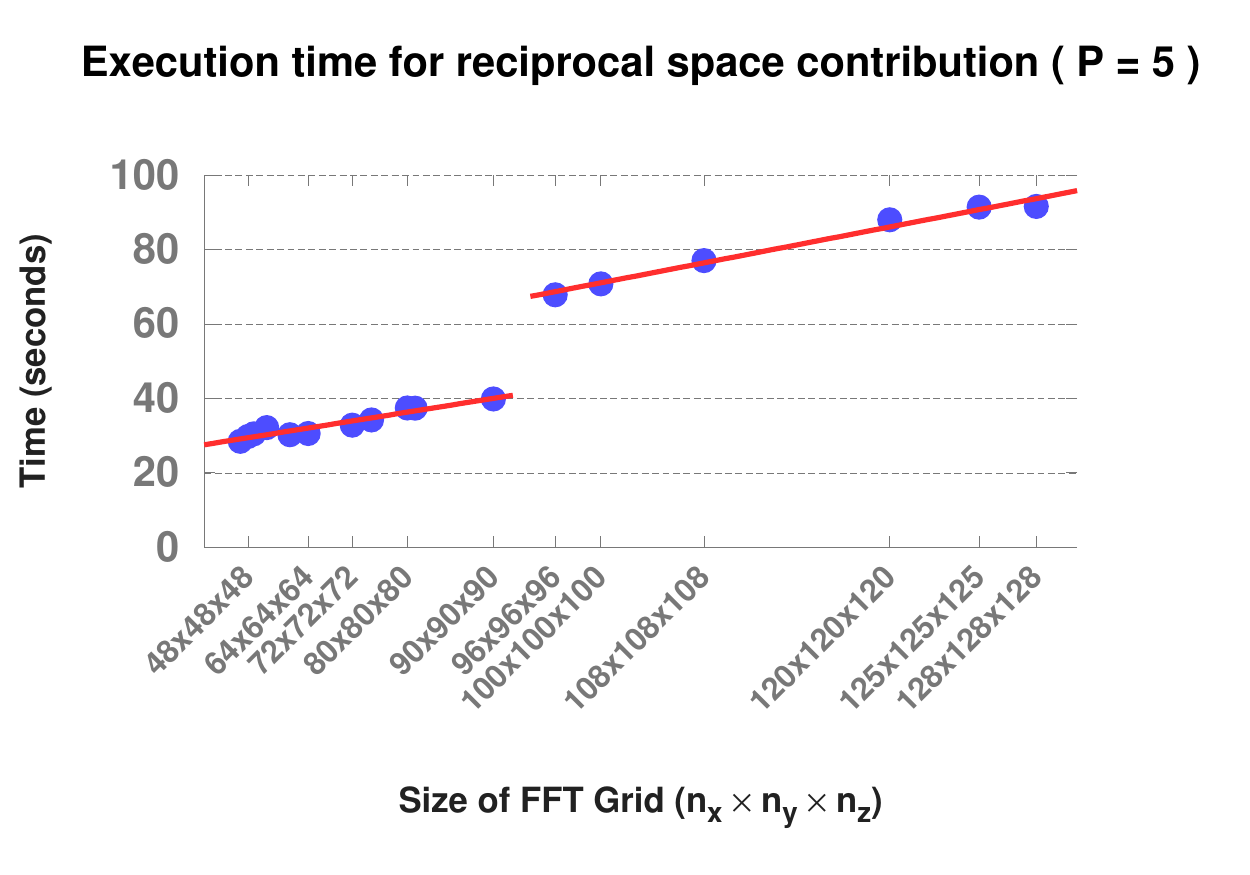}
  \caption{Samples for the execution time of the reciprocal-space contribution
  in \tcube{} (\ad{} differentiation). The data is not fitted by one single line;
  it requires two of them, with a similar slope but a shifted offset.}
  \label{fig:kspace-shift}
\end{figure}

\paragraph{Reducing the number of samples.}

Finally, the dense static sampling 
may involve a fairly large number of samples. 
For instance, \tint{} (\ik{} differentiation)
requires around 200 
samples. 
While tractable, the number of required samples
will be reduced with an adaptive sampling technique.

\subsection{Adaptive sampling}
\label{subsec:adaptive-sampling}

In light of the aforementioned issues, we present here an adaptive
strategy to exploring the search space, whose decisions are guided by the
characteristics of the simulation at hand.

The new strategy is built upon the algorithm sketched in Alg.~\ref{alg:adaptive}.
Given a fitting function {\tt f}, the list of possible values for the independent
variables {\tt xs} (either the cutoff for the real space or the grids for the
reciprocal space) and an error {\tt threshold}, the algorithm commences by sampling
the minimum, maximum, and midpoint values in {\tt xs}; the function {\tt f} is
then fitted to the measurements. If the relative error at any of the
sampled points is larger than the given {\tt threshold}, the algorithm proceeds
recursively for each half of {\tt xs}; it terminates otherwise. 

Next, we make use of a classic top-down algorithm for the segmentation of time 
series~\cite{segmentation}
that takes the series of samples collected by
Alg.~\ref{alg:adaptive} and creates a piece-wise function consisting of one or
more instances of {\tt f} with different parameters $a_1, a_2, ..., a_n$.  Six such
piece-wise functions will be created, one for the real space contribution, and
five for the reciprocal space (one per interpolation order). These functions
will then be used to model the execution time of each contribution given a
cutoff value (real space), an interpolation order and a grid size (reciprocal space).

\begin{center}
\begin{algorithm}
\begin{algorithmic}[1]
\Function{adaptive\_sampling}{f, threshold, xs}
    \Function{adaptive\_sampling\_rec}{i, j}
        \If{$(j-i) \le 1$}
            \State \Return
        \EndIf
        \State midpoint = (i+j) / 2
        \State timings(midpoint) = sample(midpoint)
        \State x = [xs(i), xs(midpoint), xs(j)]
        \State y = [timings(i), timings(midpoint), timings(j)]
        \If{error(f, x, y) $>$ threshold}
            \State \Call{adaptive\_sampling\_rec}{i, midpoint}
            \State \Call{adaptive\_sampling\_rec}{midpoint, j}
        \EndIf
    \EndFunction
    \State n\_xs = length(xs)
    \State timings = array(n\_xs)
    \State timings(1) = sample(1)
    \State timings(n\_xs) = sample(n\_xs)
    \State \Call{adaptive\_sampling\_rec}{1, n\_xs}

    \State \Return timings
\EndFunction
\end{algorithmic}
    \caption{{\bf: Adaptive sampling.}}
    \label{alg:adaptive}
\end{algorithm}
\end{center}

\begin{figure}[!h]
  \centering
  \includegraphics[width=0.6\textwidth]{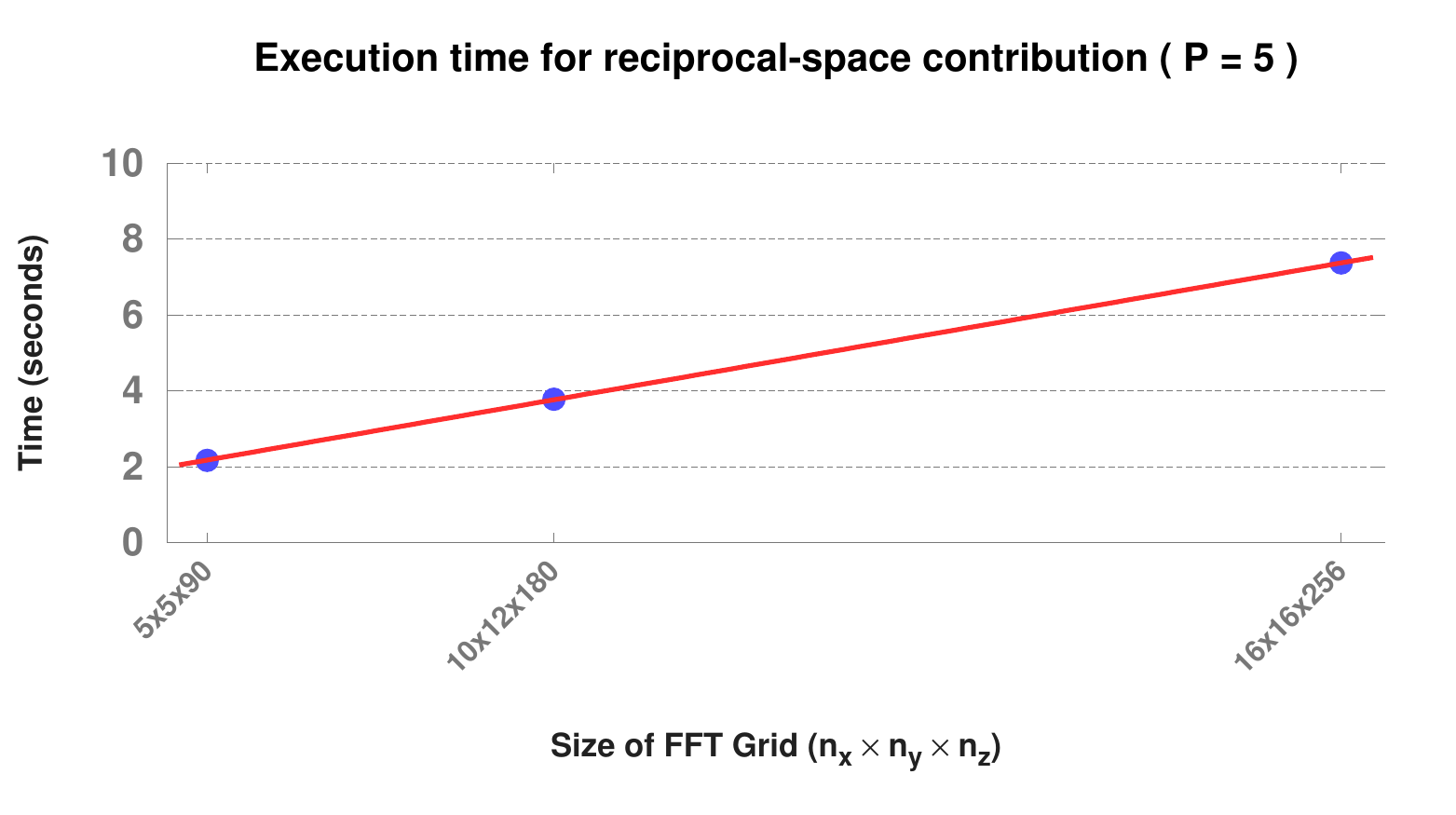}
  \caption{Samples for the execution time of the reciprocal-space contribution
  in \tint{} (\ik{} differentiation). Dynamic strategy.}
  \label{fig:kspace-si-dyn}
\end{figure}

\begin{figure}[!h]
  \centering
  \includegraphics[width=0.6\textwidth]{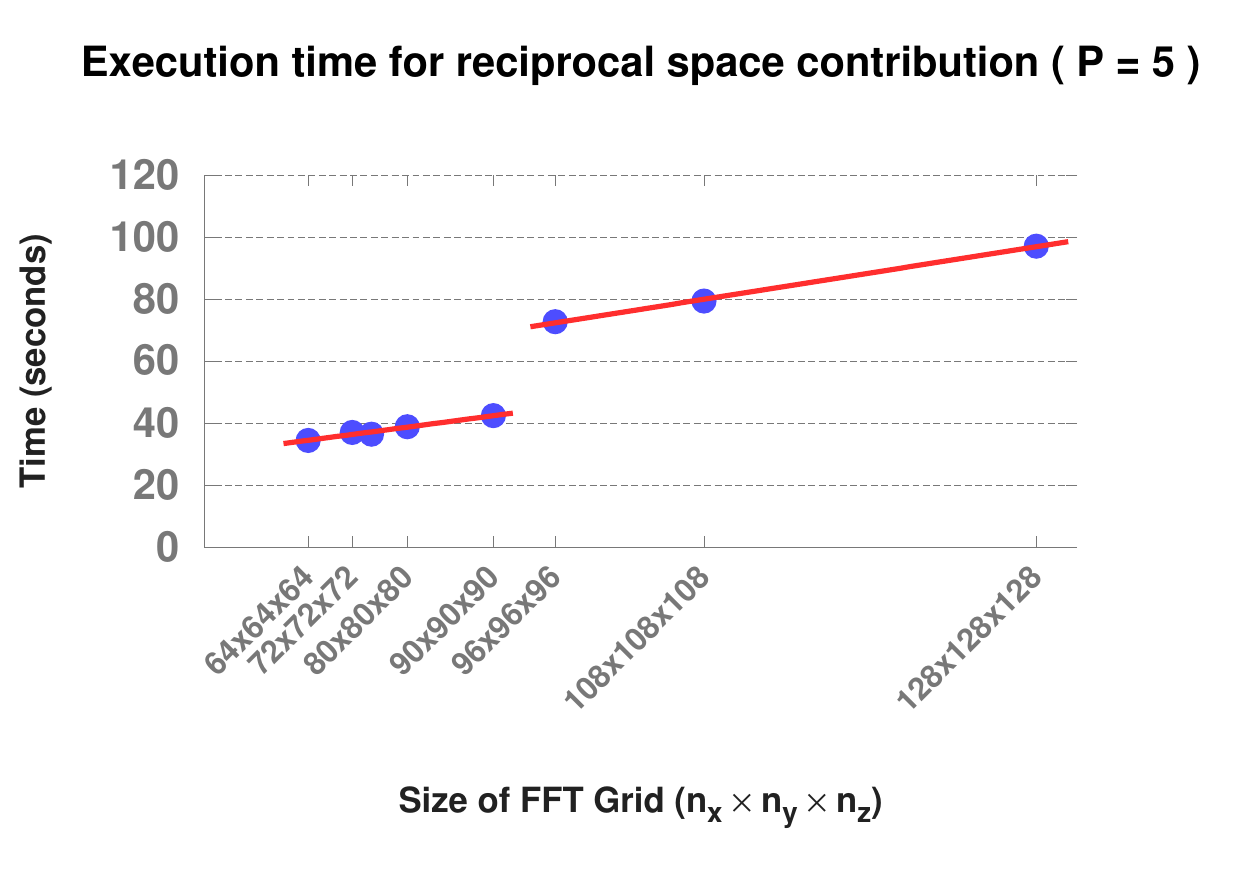}
  \caption{Samples for the execution time of the reciprocal-space contribution
  in \tcube{} (\ad{} differentiation). Dynamic strategy.}
  \label{fig:kspace-lc-dyn}
\end{figure}

\vspace{3ex}
This adaptive strategy reduces the amount of sampling. For instance,
when sampling for the reciprocal space in \tint{},
the static full sampling (Sec.~\ref{sec:static}) and the adaptive sampling 
use 18 and 3 data points per interpolation order, respectively
(see Fig.~\ref{fig:kspace-full-sampling} vs. Fig.~\ref{fig:kspace-si-dyn}).
In the ``less friendly'' \tcube{} scenario, 
17 and 8 samples per interpolation order are used, respectively
(see Fig.~\ref{fig:kspace-shift} vs. Fig.~\ref{fig:kspace-lc-dyn}).
Not only the number of samples is reduced; the shift is also correctly identified,
thus improving the accuracy of the predictions.

Finally, the effects of the cutoff in the computation of the reciprocal-space
term are addressed as follows: instead of fixing the value of the cutoff to
2.0$\sigma$, we sample for the minimum and the maximum values of the cutoff
present in the accurate subspace, and interpolate for intermediate values.
To compensate for the increase in number of samplings, we further adjust 
the sampling to the target system. 
Concretely, since we are only interested
in accurate enough configurations, we only sample the interpolation orders present in the 
accurate subspace configurations. Likewise, we adjust the range of grid sizes for
the sampling of the reciprocal space and the range of cutoffs for
the sampling of the real space.

As a result of the described adaptive sampling, we obtain accurate predictions
at a reduced sampling cost.  In Tab.~\ref{tab:si-dyn}, we present results for
\tint{}  (\ik{} differentiation). 
While the static sampling required 99 samples and attained a relative error
of 4.98\%, the dynamic strategy only required 37 samples, and achieved a
reduced relative error of 2.46\%.
Most importantly, the dynamic approach still selects the fastest configuration as the
optimal choice.

\begin{table}[!h]
  \centering
  \begin{tabular}{c @{\hspace*{1em}} c @{\hspace*{1em}}|@{\hspace*{1em}} c c @{\hspace*{1em}}|| 
     @{\hspace*{1em}} c @{\hspace*{1em}}c @{\hspace*{1em}}c @{\hspace*{1em}}c} \toprule
      \multicolumn{2}{c| @{\hspace*{1em}}}{\bf Empirical} & 
      \multicolumn{2}{c||@{\hspace*{1em}}}{\bf Prediction} &
      \multicolumn{4}{c}{\bf Configuration} \\ \midrule
      Ranking & Time & Time & Error & \ewald{} & \rc{} & \p{} & \gs{} \\ \midrule
      \#1 & 8.770s & 8.663s & (-1.23\%) & 0.56\sunits{} & 4.60$\sigma$ & 4 & $10 \times 10 \times 160$ \\
      \#2 & 8.920s & 8.808s & (-1.26\%) & 0.60\sunits{} & 4.40$\sigma$ & 4 & $12 \times 12 \times 180$ \\
      \#3 & 9.023s & 8.919s & (-1.16\%) & 0.63\sunits{} & 4.30$\sigma$ & 4 & $12 \times 12 \times 216$ \\
      \#4 & 9.119s & 9.024s & (-1.06\%) & 0.58\sunits{} & 4.50$\sigma$ & 5 & $10 \times 10 \times 160$ \\
      \#5 & 9.299s & 8.886s & (-4.64\%) & 0.64\sunits{} & 4.20$\sigma$ & 5 & $12 \times 12 \times 180$ \\ \bottomrule
  \end{tabular}
  \caption{Predictions for \tint{} (\ik{} differentiation) based on dynamic sampling.}
  \label{tab:si-dyn}
\end{table}

Table~\ref{tab:lc-dyn} collects similar results for the \tcube{}
scenario (\ad{} differentiation). The average relative error is of 3.72\%,
obtained with 51 samples. As in the previous example, our methodology 
again selects the fastest configuration as optimal.

\begin{table}[!h]
  \centering
  \begin{tabular}{c @{\hspace*{1em}} c @{\hspace*{1em}}|@{\hspace*{1em}} c c @{\hspace*{1em}}|| 
     @{\hspace*{1em}} c @{\hspace*{1em}}c @{\hspace*{1em}}c @{\hspace*{1em}}c} \toprule
      \multicolumn{2}{c| @{\hspace*{1em}}}{\bf Empirical} & 
      \multicolumn{2}{c||@{\hspace*{1em}}}{\bf Prediction} &
      \multicolumn{4}{c}{\bf Configuration} \\ \midrule
      Ranking & Time & Time & Error & \ewald{} & \rc{} & \p{} & \gs{} \\ \midrule
      \#1 & 91.663s & 94.163s & (+2.65\%) & 0.52\sunits{} & 5.30$\sigma$ & 6 & $90 \times 90 \times 90$ \\
      \#2 & 92.356s & 96.552s & (+4.35\%) & 0.50\sunits{} & 5.50$\sigma$ & 6 & $80 \times 80 \times 80$ \\
      \#3 & 92.374s & 95.707s & (+3.48\%) & 0.49\sunits{} & 5.50$\sigma$ & 5 & $90 \times 90 \times 90$ \\
      \#4 & 93.417s & 98.256s & (+4.93\%) & 0.48\sunits{} & 5.60$\sigma$ & 6 & $75 \times 75 \times 75$ \\
      \#5 & 93.675s & 98.289s & (+4.69\%) & 0.47\sunits{} & 5.70$\sigma$ & 5 & $80 \times 80 \times 80$ \\ \bottomrule
  \end{tabular}
  \caption{Predictions for the \tcube{} scenario (\ad{} differentiation) based on dynamic sampling.}
  \label{tab:lc-dyn}
\end{table}

\section{Experimental results}
\label{sec:experiments}
Through a number of case studies, 
we now discuss in detail the practical
benefits of our methodology to speed up scientific codes.
For each experiment, we report the time required by our tool
to estimate the best selection of parameters,
and the speedup with respect to the parameters chosen by an expert.
Moreover, we quantify the benefits brought to the end user
in terms of amount of additional research enabled, that is, 
for a characteristic setting where a scientist is granted
10 million core-hours at a supercomputer, and each simulation
is to be run for 50 million timesteps, we calculate how many
additional simulations are now possible.

\subsection{Experimental setup}

The experiments were run on two computing clusters.
The first one, referred as {\em Harpertown}, consists of 32 nodes;
each node comprises two four-core Intel Harpertown E5450 processors, 
operating at a frequency of 3.00GHz, and is equipped with 16GB of RAM.
The second cluster is the {\em SuperMUC} supercomputer at the Leibniz
Supercomputing Center; each node of SuperMUC consists of 16-core nodes 
based on Sandy Bridge-EP Xeon E5-2680 processors, operating at
a frequency of 2.7GHz. Each node is equipped with 32GB of RAM.
In all cases, the simulations were run using LAMMPS (version 22Jan14), FFTW, 
and OpenMPI. 

In all cases, Lennard Jones particles with energy $\epsilon$ and diameter
$\sigma$ were randomly placed in the domain (or box for interfacial systems).
Then, the systems were equilibrated for 100{,}000 timesteps after minimization
using soft potential.  The simulations were run at a temperature of
0.7$\epsilon$/$k_B$ using a Nos\'e-Hoover thermostat~\cite{NoseHoover} with
damping factor of 10$\tau$.\footnote{Here $\epsilon$ is the depth of the
Lennard-Jones potential and $k_B$ the Boltzmann constant.}

For the type of systems used in our experiments, the developer of
the LAMMPS PPPM solver for dispersion recommends to set the target 
accuracy to 0.001$\epsilon/\sigma$ for the real space and 0.1$\epsilon/\sigma$ 
for the reciprocal space. As for a fair comparison against the 
parameters automatically selected by our tool, he suggests to
set a cutoff value of $r_c = 3.0\sigma$ and let the solver set the other
parameters (Ewald parameter, interpolation order, and grid size).

\subsection{Case Study 1: Bulk system}
\label{subsec:exp-bulk}
 As a first case study,
 we consider a bulk system consisting of
 256{,}000 Lennard-Jones (LJ) particles randomly placed in a domain of length
 $55\sigma \times 55\sigma \times 110\sigma$.
 The computations were carried out on 12 Harpertown
 nodes (i.e., 96 processors). 

 To determine the benefit of our methodology for automatic parameter selection,
 we compare it with a human expert's best guess, and with the empirical fastest. 
 The execution time for a sample of 1{,}000 timesteps with the configuration
 suggested by the developer is collected in
 Tab.~\ref{tab:bulk-results}, rows 2 and 4. The empirical fastest configurations
 and the corresponding timings are displayed in rows 3 and 5.
 Next, we allowed our prediction tool to run the necessary samples
 to estimate the execution time of the parameterizations at the
 frontier configurations, and select the predicted fastest.
 The sampling and prediction took about four hours.
 Table~\ref{tab:bulk-results} collects the results for our tool.
 In both cases, the predictions match the best empirically-determined
 configurations. 
 The final choice of our tool is to use \ad{} differentiation with the following
 parameters: (\ewald{}$\; =0.84$, \rc{}$\; =3.30\sigma$, \p{}$\; =4$, 
 \gs{}$\; =45\times45\times90$). The choice not only coincides with the empirically
 fastest, but it also reduces execution time by 35\% with respect to the developer's best
 guess.

 \begin{table}[!h]
     \centering
     \begin{tabular}{l @{\hspace*{1em}}||@{\hspace*{1em}} c @{\hspace*{1em}} c @{\hspace*{1em}} c @{\hspace*{1em}} c @{\hspace*{1em}} c @{\hspace*{1em}} c} \toprule
         {\bf Approach} & {\bf Diff.} & {\bf Time}  &     {\bf \ewald{}} & {\bf \rc{}} & {\bf \p{}} & {\bf \gs{}} \\\midrule
         Expert guess   & \ik{} & 69.36s & 0.965\sunits{} &         3.00$\sigma$ &          5    & $64 \times 64 \times 125$ \\
         Empirical      & \ik{} & 44.39s & 0.810\sunits{} &         3.40$\sigma$ &          4    & $40 \times 40 \times 81\phantom{a}$ \\\midrule
         Expert guess   & \ad{} & 61.87s & 0.965\sunits{} &         3.00$\sigma$ &          5    & $60 \times 60 \times 120$ \\
         Empirical      & \ad{} & 40.16s & 0.840\sunits{} &         3.30$\sigma$ &          4    & $45 \times 45 \times 90\phantom{a}$ \\\midrule
         Prediction     & \ik{} & 44.39s & 0.810\sunits{} &         3.40$\sigma$ &          4    & $40 \times 40 \times 81\phantom{a}$ \\
         Prediction     & \ad{} & 40.16s & 0.840\sunits{} &         3.30$\sigma$ &          4    & $45 \times 45 \times 90\phantom{a}$ \\\bottomrule
     \end{tabular}
     \caption{Results for the {\em Bulk system}. Expert choices and best empirical configurations 
         for each differentiation mode. The predicted best configurations
         coincide with those that empirically proved to be fastest.
       }
     \label{tab:bulk-results}
 \end{table}

 %
 %

 Given the characteristic scenario in Sec.~\ref{sec:experiments}, with the choice of
 parameters of the expert, one may run up to 121 simulations, while the
 automatically chosen parameters enable 187 of them; that is, the scientist
 can carry out a 50\% more research.

\subsection{Case study 2: Interfacial system}
 As a second case study, we consider an interfacial system consisting of
 128{,}000 LJ particles randomly placed in a box of length
 $55\sigma \times 55\sigma \times 55\sigma$, 
 centered in a domain of length
 $55\sigma \times 55\sigma \times 165\sigma$.
 The computations were carried out on 6 of the Harpertown nodes,
 that is, 48 processes in total.

As in the previous example, we first run the experiments based
on the developer recommended configurations, shown
in rows 1 and 3 of Tab.~\ref{tab:int-results}. 
Then, we timed the configurations in the frontier (rows 2 and 4).
Finally, we ran our tool (for about three hours), which selected the configurations in rows 5 and 6.
In this case, our tool selected the best configuration for the \ik{}
differentiation; as for the \ad{} case, while it did not find the absolute best 
configuration, its choice is less than 2\% away from the optimal. 
Most importantly, when compared to the expert's guess, the automatically
selected parameters yield a reduction in the execution time of 13\% and 27\%
for the \ad{} and \ik{} differentiations, respectively.

 \begin{table}[!h]
     \centering
     \begin{tabular}{l @{\hspace*{1em}}||@{\hspace*{1em}} c @{\hspace*{1em}} c @{\hspace*{1em}} c @{\hspace*{1em}} c @{\hspace*{1em}} c @{\hspace*{1em}} c} \toprule
         {\bf Approach} & {\bf Diff.} & {\bf Time}       & {\bf \ewald{}} & {\bf \rc{}} & {\bf \p{}} & {\bf \gs{}} \\\midrule
         Expert guess   & \ik{} & 60.78s & 0.92\sunits{} & 3.00$\sigma$ & 5 & $48 \times 48 \times 144$ \\
         Empirical      & \ik{} & 44.52s & 0.77\sunits{} & 3.50$\sigma$ & 2 & $32 \times 32 \times 96\phantom{0}$ \\\midrule
         Expert guess   & \ad{} & 46.62s & 0.92\sunits{} & 3.00$\sigma$ & 5 & $45 \times 45 \times 144$ \\
         Empirical      & \ad{} & 39.79s & 0.81\sunits{} & 3.40$\sigma$ & 3 & $32 \times 36 \times 100$ \\\midrule
         Prediction     & \ik{} & 44.52s & 0.77\sunits{} & 3.50$\sigma$ & 2 & $32 \times 32 \times 96\phantom{0}$ \\
         Prediction     & \ad{} & 40.55s & 0.74\sunits{} & 3.60$\sigma$ & 2 & $30 \times 30 \times 90\phantom{0}$ \\\bottomrule
     \end{tabular}
     \caption{Results for the {\em Interfacial system}. Expert choices and best empirical configurations 
         for each differentiation mode. The predicted best configuration is
         extremely close to the one that empirically proved to be the fastest one,
         and is considerably faster than the expert's choice.
       }
     \label{tab:int-results}
 \end{table}

In reference to the characteristic setting outlined in Sec.~\ref{sec:experiments}, with the choice of
parameters of the developer, one may run up to 320 simulations, while the
automatically chosen parameters enable 370 of them; that is, the scientist
can carry out a 16\% more research.

 %
 %

\subsection{Case study 3: Large Interfacial system}

We turn now our attention to larger simulations requiring at least
hundreds of processors. Our third case study consists of an interfacial
system including 512{,}000 particles placed in a box of length 
$64\sigma \times 64\sigma \times 128\sigma$, centered in a domain of length
$64\sigma \times 64\sigma \times 256\sigma$.
The computations were carried out on 32 Harpertown
nodes (i.e., 256 processors). 

Table~\ref{tab:intl-results} collects the timings for the configurations
selected by the expert user, as well as the configuration automatically
selected by our tool. Since we already demonstrated the accuracy of
our predictions, and with the purpose of limiting the usage of resources in the experiment,
we do not run the simulation for each of the configurations in the frontier.
The 
automatically
selected parameters attain a remarkable speedup with respect to the developer's best guess of 2.33x.

 \begin{table}[!h]
     \centering
     \begin{tabular}{l @{\hspace*{1em}}||@{\hspace*{1em}} c @{\hspace*{1em}} c @{\hspace*{1em}} c @{\hspace*{1em}} c @{\hspace*{1em}} c @{\hspace*{1em}} c} \toprule
         {\bf Approach} & {\bf Diff.} & {\bf Time}  &     {\bf \ewald{}} & {\bf \rc{}} & {\bf \p{}} & {\bf \gs{}} \\\midrule
         Expert guess   & \ik{} & 165.5s & 0.947\sunits{} & 3.00$\sigma$ & 5 & $64 \times 64 \times 256$ \\
         Expert guess   & \ad{} & 146.1s & 0.947\sunits{} & 3.00$\sigma$ & 5 & $64 \times 64 \times 256$ \\
         Prediction     & \ad{} & \phantom{0}62.8s & 0.850\sunits{} & 3.30$\sigma$ & 3 & $54 \times 54 \times 216$ \\\bottomrule
     \end{tabular}
     \caption{Results for the {\em Large Interfacial system}. 
         Expert choices for each differentiation mode and the predicted best configuration.
         The automatically selected parameters attain a speedup of 2.33x with respect to
         the best expert guess.
       }
     \label{tab:intl-results}
 \end{table}

In terms of the aforementioned characteristic scenario, instead of only 19 simulations,
the user can now run 45 of them.

\subsection{SuperMUC: Different workloads on a supercomputer}

In this final case study we take a slightly different direction
to demonstrate the applicability of our prototype when the target architecture
is a supercomputer. 
To this end, we select an interfacial system with 2 million particles 
placed in a box of size
$128\sigma \times 128\sigma \times 128\sigma$,
centered within a domain of size
$128\sigma \times 128\sigma \times 256\sigma$.
As in the previous examples, the desired accuracy is set to
0.001$\epsilon/\sigma$ for the real space, and 0.1$\epsilon/\sigma$ for the reciprocal space.
Now, to confer more breadth to the study, we consider simulation
runs on different number of cores: 512, 1024, and 2048, and
thus with different workloads per core.

Due to the limited resources at our disposal to carry out
this experiment, we only consider the \ad{} differentiation.
In Tab.~\ref{tab:supermuc} we collect the results
for 1{,}000 timesteps of the simulation using the developer's
suggestion (column 2) and the configuration selected by our
tool (column 3). 
In all three cases, the expert configuration is:
\ewald{} $ = 0.947$, cutoff = 3.0, interpolation order = 5, and
grid size = ($125 \times 125 \times 243$), while our tool selects:
\ewald{} $ = 0.85$, cutoff = 3.3, interpolation order = 3, and
grid size = ($108 \times 108 \times 216$).
Irrespective of the workload, as long as it reaches
a reasonable minimum of 1000 particles per processor, the automatic
selected parameters achieve speedups between a 11\% and
a 16\%.

 \begin{table}[!h]
     \centering
     \begin{tabular}{c @{\hspace*{1em}} ||
                       @{\hspace*{1em}} r
                       @{\hspace*{1em}} r
                       @{\hspace*{1em}} c} \toprule
         {\bf \# procs} & 
         \multicolumn{1}{@{\hspace*{1em}}c@{\hspace*{1em}}}{\bf Expert} & 
         \multicolumn{1}{c@{\hspace*{1em}}}{\bf Prediction} &
         {\bf Speedup} \\\midrule
         2048            &  8.19 secs. &  7.34 secs. & 11.6\% \\
         1024            & 13.16 secs. & 11.36 secs. & 15.8\% \\
         \phantom{0}512  & 23.96 secs. & 20.55 secs. & 16.6\% \\\bottomrule
     \end{tabular}
     \caption{Results for the experiments in SuperMUC. Independently of the 
         workload per processor, the automatically selected parameters attain
         speedups between 11\% and 16\% with respect to the expert suggestion.
     }
     \label{tab:supermuc}
 \end{table}

Given 10 million core-hours granted, a scientist can
now run 48, 62, and 68 50-million timestep simulations,
instead of 43, 53, and 59, using 2048, 1024, and 512
processors, respectively.

\section{Conclusions}

We presented a methodology for the automatic selection of parameters for
simulation software governed by parameters that affect performance and/or
accuracy. When using such software, users face a challenging trade-off problem
between accuracy and performance, and even the expert practitioners and the
actual developers spend considerable effort and time to find relatively good
choices for the parameters.
We developed a tool implementing the methodology for the \PD{} solver
for dispersion interactions from the LAMMPS suite,
which not only spares the user from spending valuable time on trial and
error evaluation, but also finds close-to-optimal configurations that
attain the requested accuracy with close-to-minimum execution time.

The methodology proceeds in three steps. In the first step,
the parameters of interest are identified, the continuous ones
are discretized, and acceptable ranges of values are given
for each of them. The outcome of this step is a search
space $\mathcal{S}$. In the second step, the methodology splits $\mathcal{S}$ 
into accurate and inaccurate subspaces ($\mathcal{S_A}$
and $\mathcal{S_I}$), according to the accuracy requested by the user;
only $\mathcal{S_A}$ is further considered.
In the last step, a reduced number of samples (timings) are taken and fitted to
simple functions based on the asymptotic computational
cost of the method under study. These functions are then used to model the
performance of each configuration in the frontier $\mathcal{F}$
(the accurate configurations in the boundary between $\mathcal{S_A}$ and
$\mathcal{S_I}$) and to select the estimated fastest one.

We showed that in order to accurately predict performance and to
find close-to-optimal configurations,
it is crucial to deeply understand the accuracy and performance
behavior of the method. Further, the structure of the problem
may be exploited to reduce the complexity of the search, for instance,
by splitting the search in a divide and conquer fashion, and to
speed up the search process.

The application of our prototype, which completes the search in at most a few
hours, is much faster than manual trial and error of many different
configurations, and finds close-to-optimal configurations that achieve speedups
with respect to the best expert guesses ranging from 1.10x to 2.33x. The
corresponding reduction of time-to-solution allows the practitioners to perform
much more research, that is, run many more simulations, within a given core-hour
budget, allowing them to gain deeper insight in their investigations.

\begin{acknowledgements}

The authors gratefully acknowledge financial support from the Deutsche
Forschungsgemeinschaft (German Research Association) through grant GSC 111,
and from the Hans Hermann Voss-Stiftung through grant  OPSF224--ADAMS.
We thank RWTH Aachen IT Center and the Gauss Centre for Supercomputing/Leibniz
Supercomputing  Centre (project ID: pr94te) for providing the necessary computing resources,
as well as Daniel Tameling and Rolf Isele-Holder for their support
on the usage of LAMMPS and the PPPM method.

\end{acknowledgements}


\newpage

\appendix
\section{Test cases for the preliminary study of the methodology}
\label{app:scenarios}

\begin{figure}[!h]
    \begin{center}

        \begin{subfigure}[b]{0.2\textwidth}
            \centering
            \includegraphics[height=2cm]{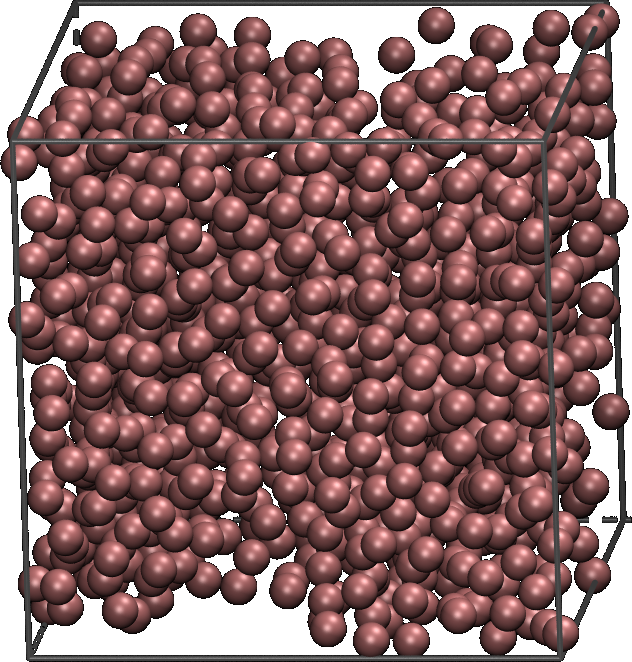}
            \caption{}
            \label{fig:mc}
        \end{subfigure}
%
%
        \begin{subfigure}[b]{0.3\textwidth}
            \centering
            \includegraphics[height=.5cm]{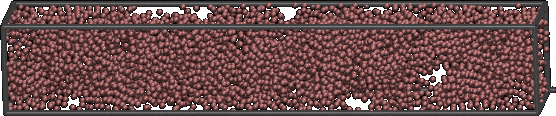}
            \caption{}
            \label{fig:sb}
        \end{subfigure}
        \begin{subfigure}[b]{0.4\textwidth}
            \centering
            \includegraphics[height=.5cm]{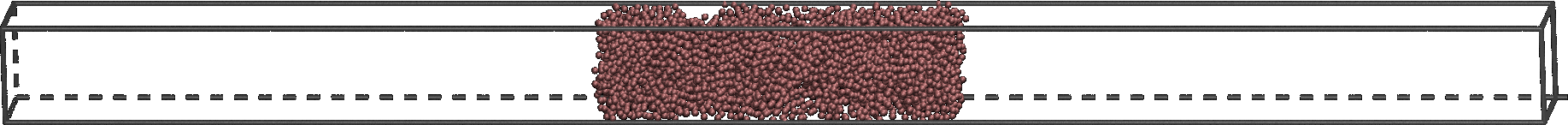}
            \caption{}
            \label{fig:si}
        \end{subfigure}
    \end{center}
    \caption{Illustration of the domain and initial position of the particles for the 
    test cases collected in Tab.~\ref{tab:scenarios}. a) Large cube, b) Small bulk, c) Small interface.}
    \label{fig:boxes}
\end{figure}

In this appendix we collect a number of test cases we used for the preliminary
study of our methodology and the development of the prototype.
Table~\ref{tab:scenarios} provides, for each test, the size of the domain, the
number of Lennard-Jones particles (with energy $\epsilon$ and diameter
$\sigma$) in the system, the number of cores used to run the experiments, and
the target accuracy (as a single value for the combined root mean square). 
In all cases, the particles were randomly placed in the domain (in a centered
box for the interfacial system, see Fig.~\ref{fig:si}). Then, the systems were
equilibrated for 100{,}000 timesteps after minimization using soft potential.
The simulations were run at a temperature of 0.7$\epsilon$/$k_B$ using a
Nos\'e-Hoover thermostat~\cite{NoseHoover} with damping factor of
10$\tau$.\footnote{Here $\epsilon$ is the depth of the Lennard-Jones potential
and $k_B$ the Boltzmann constant.}

\begin{table}[!h]
    \centering
    \begin{tabular}{l @{\hspace{5mm}} l @{\hspace{5mm}} r @{\hspace{5mm}} c @{\hspace{5mm}} c} \toprule
        {\bf Scenario} &
        \multicolumn{1}{c}{\bf Domain size [$\sigma{}^3$]} & 
        \multicolumn{1}{c@{\hspace{5mm}} }{\bf \# particles} &
        {\bf \# processes} &
        {\bf Accuracy [$\epsilon/\sigma$]} \\ \midrule
        Large Cube  & $88.08 \times 88.08 \times 88.08$ & 512{,}000 & 96 & $10^{-4}$ \\
        Small Bulk  & $11.01 \times 11.01 \times 66.06$ & 6{,}000 & 8 & $10^{-4}$ \\
        Small Interface & $11.01 \times 11.01 \times 176.16$ & 4{,}000 & 
                         8 & $5 \times 10^{-4}$ \\ \bottomrule
    \end{tabular}
    \caption{Selection of test cases used in the study and development of the presented
    methodology and tool.}
    \label{tab:scenarios}
\end{table}

In Fig.~\ref{fig:boxes} we illustrate the shape and
distribution of the particles in each case.  In the {\em Small Interface}
scenario, the particles are placed in a centered box of size $11.01 \times
11.01 \times 44.04$.

\end{document}